\documentclass{aa}  

\usepackage{textcomp}
\usepackage{lastpage}
\usepackage{comment}
\usepackage{csquotes}
\usepackage{gensymb}
\usepackage{float}
\usepackage{hyperref}
\usepackage{graphicx}
\usepackage{placeins}
\usepackage{xcolor}
\usepackage{upgreek}
\usepackage{txfonts}
\usepackage{outlines}
\usepackage[separate-uncertainty=true,multi-part-units=single]{siunitx}

\DeclareSIPrefix\micro{\upmu}{-6}

\DeclareSIUnit{\erg}{erg}
\DeclareSIUnit{\parsec}{pc}
\DeclareSIUnit{\jansky}{Jy}
\DeclareSIUnit{\beam}{beam}
\DeclareSIUnit{\asec}{arcsec}

\begin{document}

   \title{Investigating the radio emission in the Perseus cluster with \\ LOFAR sub-80~MHz LBA observations}

   \subtitle{}

   \author{C. Groeneveld \inst{\ref{ira},\ref{strw}}\thanks{E-mail: \url{c.groeneveld@ira.inaf.it}} \and R.~J. van Weeren \inst{\ref{strw}} \and M.-L. Gendron-Marsolais \inst{\ref{laval}} \and E. Osinga \inst{\ref{dunlap}} \and A. Botteon \inst{\ref{ira}} \and F. de Gasperin\inst{\ref{ira}} \and M. Cianfaglione \inst{\ref{ira},\ref{unibo}} \and G. di Gennaro \inst{\ref{ira}} \and G. Brunetti \inst{\ref{ira}} \and R. Cassano \inst{\ref{ira}}}

    \institute{%
    INAF -- Istituto di Radioastromia, via P. Gobetti 101, 40129 Bologna, Italy \label{ira}
    \and Leiden Observatory, Leiden University, Einsteinweg 55, 2333 CC Leiden, the Netherlands \label{strw}
    \and Département de physique, de génie physique et d’optique, Université Laval, Québec, QC G1V 0A6, Canada \label{laval}
    \and Dunlap Institute for Astronomy and Astrophysics, University of Toronto, 50 St. George St, Toronto, ON M5S 3HM5S 3H4, Canada\label{dunlap}
    \and Dipartimento di Fisica ed Astronomia, Università di Bologna, Via Gobetti 93/2, 40129 Bologna, Italy \label{unibo}}
    
   \date{Received 16 January 2026, Accepted 10 March 2026}

  \abstract{The Perseus cluster is a nearby cool-core galaxy cluster that hosts an archetypal radio mini-halo. Recent Low Frequency Array (LOFAR) High Band Antenna (HBA) observations at $120-168$~MHz have revealed the presence of a giant radio halo within the cluster with a size of 1.1 Mpc enveloping the mini-halo.
  By exploring the spectral properties of the radio emission at low frequencies, we can gain deeper insights into the nature of this emission and improve our understanding of its origin.
  Here we present LOFAR Low Band Antenna (LBA) images of the cluster between $30.0-57.7$ MHz, with a resolution of $19.2\arcsec \times 15.0\arcsec$ and a r.m.s. noise of 3.7 mJy~beam$^{-1}$. In our images, we detect both the mini-halo and giant radio halo.
  We measured the spectral indices between 44 and 144~MHz of the mini-halo and giant radio halo to be $-1.34\pm 0.10$, and $-1.01\pm 0.11$, respectively. An alternative and more direct measurement of the spectrum of the giant radio halo results in a spectral index of $-1.28\pm 0.15$. The discrepancy between both values is caused by the poor ionospheric conditions.

  In addition, we study two X-ray `ghost cavities' in the cluster. These cavities are thought to have been produced by an older outburst from the central AGN 3C\,84. We measure a spectral index between 44 and 144~MHz for the radio plasma in these cavities of $-1.86 \pm 0.12$ and $-1.90 \pm 0.12$ for the northwest and southern ghost cavities, respectively. Furthermore, by including VLA 352~MHz data, we find that the spectrum steepens at higher frequencies. These results are consistent with the ghost cavities being filled with old and aged radio plasma.
  We also detect the tailed radio galaxies NGC~1265 and IC~310.
  In our analysis, these sources show signs of spectral steepening along their tails.

  }


   \maketitle

\section{Introduction}
\label{sec:intro}
The Perseus galaxy cluster is a nearby \citep[$z=0.01767$;][]{2020A&A...633A..42S} cool-core galaxy cluster. 
Its proximity and thus its large angular size and brightness make it one of the most studied galaxy clusters across many different wavelength ranges, from radio \citep[e.g.][]{1966Natur.210Q..80B,1983SvAL....9..305S,2011A&A...526A...9B,2017MNRAS.469.3872G,2020MNRAS.499.5791G} and optical \citep[e.g.][]{2025A&A...697A..13K,2017ApJ...839..102H} to X-rays \citep[e.g.][]{2000MNRAS.318L..65F,2006MNRAS.366..417F,2011MNRAS.418.2154F} and gamma rays \citep[e.g.][]{2006ApJ...644..148P,2009ApJ...706L.275A,2010ApJ...710..634A}.
The Perseus galaxy cluster is a prime example of the feedback between the central active galactic nucleus (AGN) and the intracluster medium (ICM). Accretion onto the supermassive black hole in the AGN drives the ejection of material via relativistic jets, which deposit energy into the surrounding ICM and cluster core. This feedback mechanism is essential for preventing a runaway cooling flow, as the radiative cooling time in the centre of cool-core clusters is shorter than the Hubble time \citep{2012ARA&A..50..455F,2012NJPh...14e5023M}.

Observationally, feedback can manifest itself as cavities in the ICM, which are revealed by high-resolution X-ray observations \citep[e.g.][]{1993MNRAS.264L..25B,1994MNRAS.270..173C,2000A&A...356..788C,2002MNRAS.331..369F}.
These cavities often coincide with the relativistic plasma originating from the radio AGN. 
This relativistic plasma can then excavate depressions in the ICM that subsequently rise buoyantly \citep{2007ARA&A..45..117M}.
Some X-ray cavities do not appear to be filled with radio-emitting plasma, and they are referred to as `ghost cavities'.
The idea is that these cavities are older and contain relativistic plasma from previous AGN outbursts that do not emit at high radio frequencies because the high-energy cosmic rays have aged through synchrotron and inverse Compton energy losses. Indeed, radio emission is detected in some of these cavities only through low-frequency radio observations \citep[e.g.][]{2002MNRAS.331..369F,2001ApJ...562L.149M,2005ApJ...625..748C,2007ApJ...659.1153W,2017A&A...605A..48K,2020MNRAS.496.2613B}.

The Perseus cluster hosts two pronounced X-ray cavities associated with the radio AGN 3C\,84 from the brightest cluster galaxy NGC\,1275 \citep[e.g.][]{1981ApJ...248...55B,1981ApJ...248...47F,1993MNRAS.264L..25B,2000A&A...356..788C}. These two cavities are coincident with the radio lobes of the central AGN.
In addition, two ghost cavities have been detected at a larger distance from the central AGN \citep[e.g.][]{2006MNRAS.366..417F}. At 74\,MHz, the Very Large Array (VLA), however, showed the presence of steep-spectrum radio emission in the direction of these cavities \citep{2002IAUS..199..189B}.
Further depressions in the X-ray emission have also been tentatively identified \citep{2011MNRAS.418.2154F}.

In addition, the Perseus cluster hosts a so-called radio mini-halo \citep[e.g.][]{1975A&A....45..223M,1982Natur.299..597N,1983SvAL....9..305S,1990MNRAS.246..477P,1992ApJ...388L..49B,2002A&A...386..456G,2017MNRAS.469.3872G}. Mini-halos are extended radio sources found in the centre of some relaxed clusters \citep{2012A&ARv..20...54F,2017ApJ...841...71G}, with typical linear sizes of up to $0.2 R_{500}$\footnote{We note that $R_{500}$ denotes the radius within which the mean density is 500 times the critical density: $\rho(R_{500})=500\rho_{crit}$.} \citep{2019ApJ...880...70G}. The origin of radio mini-halos is still poorly understood. The relativistic electrons from AGN outbursts have a significantly shorter radiative cooling time compared to the time required to diffuse over a volume with a size of several hundreds of kiloparsecs.
Therefore, two models have been proposed where electrons are accelerated to highly relativistic energies in situ.
One model invokes in situ re-acceleration of relativistic seed electrons \citep{2002A&A...386..456G,2004A&A...417....1G} by sloshing-driven turbulence, and the other model assumes that the emission is produced by secondary electrons due to proton-proton collisions in the ICM \citep{2004A&A...413...17P,2024MNRAS.527.1194K}.
Simulations \citep[e.g.][]{2013ApJ...762...78Z} have shown that turbulence generated by sloshing motions of the cool gas in the cluster core, induced by a minor cluster merger event, is sufficient to generate turbulence capable of re-accelerating electrons to the required relativistic energies.
Radio mini-halos are then confined inside cold fronts caused by the sloshing motions. 
Alternatively, it is possible that feedback from the central AGN can re-accelerate an old population of cosmic ray electrons \citep{2020MNRAS.499.2934R}.
It remains unclear what the source of the underlying seed population of low-energy (but still relativistic) electrons is, although the central AGN is a likely candidate \citep{2007ApJ...663L..61F,2008A&A...486L..31C}.  
The alternative is the `hadronic' model, where the relativistic electrons are produced as a byproduct of collisions of relativistic protons with thermal protons. 
Studying the hadronic model is challenging, but the Perseus cluster is one of the best laboratories, primarily due to the proximity of the cluster \citep{2010MNRAS.409..449P,2010ApJ...710..634A}. Recent work on gamma rays with energies of 100~GeV--TeV based on observations from the Major Atmospheric Gamma-ray Imaging Cherenkov (MAGIC) telescope has put significant constraints on the hadronic model, but the work was unable to rule it out \citep{2012A&A...541A..99A,2016A&A...589A..33A}.

Recently, \cite{2024A&A...692A..12V} presented $120-168$~MHz observations of the Perseus cluster by the Dutch LOw Frequency ARray (LOFAR), revealing diffuse emission on scales exceeding the range of the earlier detected mini-halo. This diffuse emission beyond the radio mini-halo has the characteristics of a giant radio halo. Giant radio halos \citep[for a review see][]{2019SSRv..215...16V} are typically associated with dynamically disturbed systems \citep{2001ApJ...553L..15B,2010ApJ...721L..82C}, but modern instruments are starting to unveil large-scale diffuse emission also in cool-core clusters with only a mild indication of disturbance.
In the case of the Perseus cluster, the giant radio halo was interpreted as originating from turbulence from a past off-axis merger event \citep{2025NatAs...9..925H} that preserved the cool-core.
Recent other examples of diffuse emission beyond the scale of mini-halos indicate that this extended emission has on average steeper spectra than the central mini-halo 
\citep{2017A&A...603A.125V,2021MNRAS.508.3995B,2023A&A...678A.133B,2024A&A...686A..44R}. To confirm if this also holds in the Perseus galaxy cluster, observations at lower frequencies than those presented in \cite{2024A&A...692A..12V} are crucial, as the giant halo is too faint to be detected with the Westerbork Synthesis Radio Telescope (WSRT) and VLA at higher frequencies.
Furthermore, \cite{2024A&A...692A..12V} detected radio emission from the two ghost cavities, showing that these cavities are filled with low-frequency radio-emitting plasma. To better understand the cosmic ray energetics associated with the ghost cavities, spectral studies at even lower frequencies are crucial.

In this work, we present LOFAR Low Band Antenna (LBA) observations (30.0--57.7~MHz) of the Perseus cluster.
In addition, for spectral mapping, we employed VLA P-band (225.4--478.4~MHz) data of the core of the Perseus cluster presented by \cite{2017MNRAS.469.3872G} as well as WSRT (323.9--330.1~MHz) data from \cite{2011A&A...526A...9B}.
In Sect.~\ref{sec:methods} we describe the observations and data reduction performed on the LOFAR LBA data. Section~\ref{sec:results} presents our measurements of the mini-halo and giant halo, the spectral properties of the emission in the ghost cavities, and an analysis of the tailed radio galaxies NGC 1265 and IC 310. The discussion and conclusions are provided in Sects.~\ref{sec:discussion} and~\ref{sec:conclusion}, respectively.
This paper assumes a Lambda cold dark matter cosmology ($H_0=57.7 \text{ km/s/Mpc}$; $\Omega_m=0.310$), according to \cite{2020A&A...641A...6P}.
At the redshift of the Perseus cluster ($z = 0.01767$), 1~kpc corresponds to 2.7\arcsec.

\begin{figure*}
    \centering
    \includegraphics[width=0.7\textwidth]{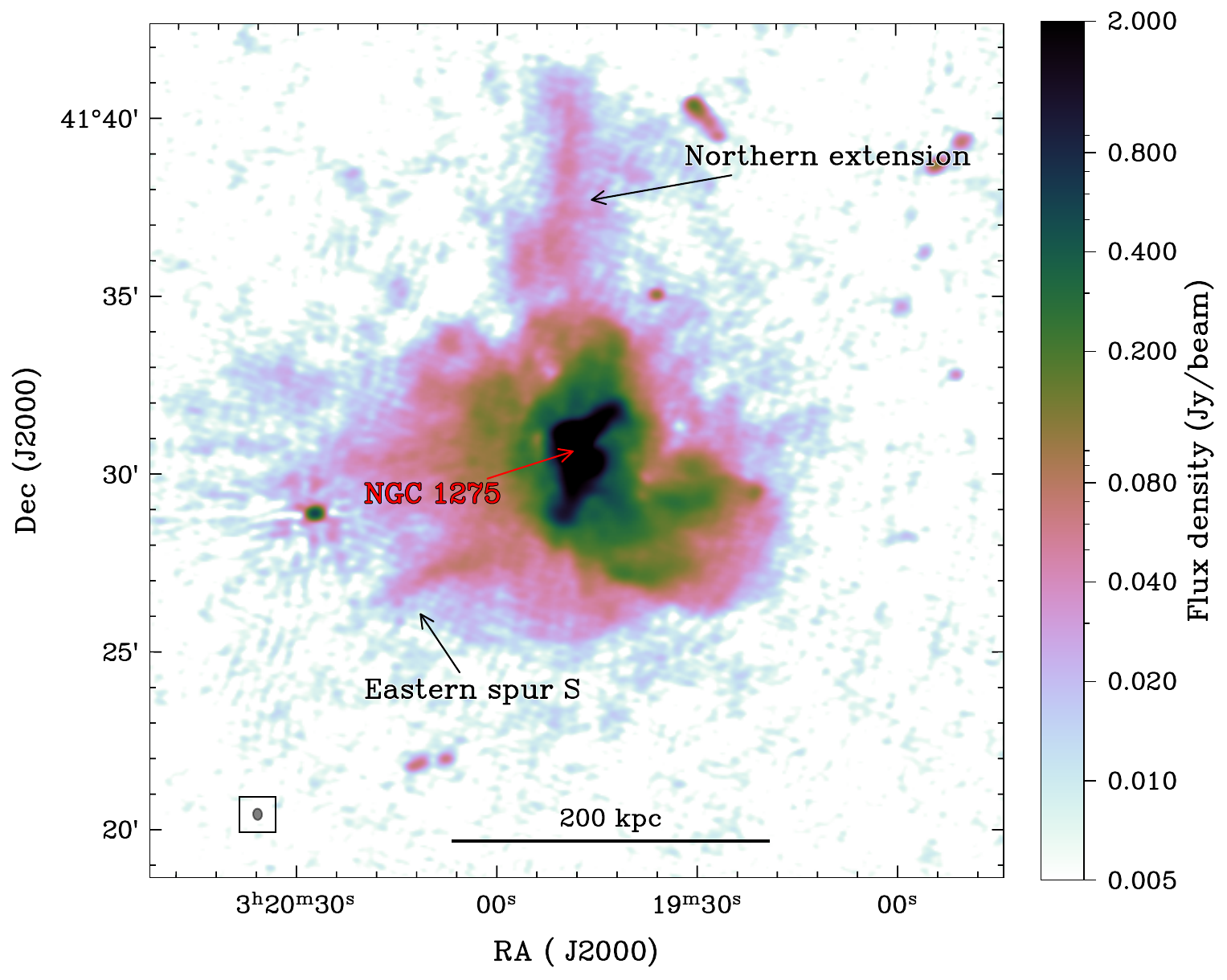}
    \caption{Full-resolution image from the LOFAR LBA (30.0--57.7~MHz of the core of the Perseus cluster. The restoring beam is shown in the bottom left (19.2\arcsec$\times$15.0
    \arcsec). The colour range is displayed on a logarithmic scale. }
    \label{fig: highres_core}
\end{figure*}

\section{Methods}
\label{sec:methods}

\begin{table*}
    \caption{Overview of the images used in this paper.}
    \label{tab:details}
    \centering
    {\fontsize{9}{7}\selectfont %
    \begin{tabular}{c|c|c|c|c|c}
         Instrument & Frequency range (MHz) & Gaussian Taper & Restoring beam & r.m.s. noise & Reference  \\ \hline
         LOFAR LBA & 30.0 -- 57.7 & -- & 19.2\arcsec $\times$ 15.0\arcsec& 2.3 mJy~beam$^{-1}$ & This work \\
         & & 60\arcsec & 67.8\arcsec $\times$ 66.5\arcsec & 7.7 mJy~beam$^{-1}$ & This work \\
         & & 120\arcsec & 145.0\arcsec $\times$ 145\arcsec & 15 mJy~beam$^{-1}$ & This work  \\ %
         LOFAR HBA & 120.2 -- 167.1 & -- &  8.2\arcsec $\times$ 4.8\arcsec  & 94.0  \textmu Jy~beam$^{-1}$ & (1) \\
         & & 60\arcsec & 90.1\arcsec $\times$ 70.2\arcsec & 823 \textmu Jy~beam$^{-1}$ & (1) \\
         & & 60\arcsec & 145\arcsec $\times$ 145\arcsec & 1.46 mJy~beam$^{-1}$ & (1) \\
         VLA P-band (A Array) & 225.4 -- 478.4 & -- & 17.7\arcsec $\times$ 15.0\arcsec & 1.5 mJy~beam$^{-1}$ & (2) \\
         WSRT & 323.9 -- 330.1 & -- & 96.0\arcsec $\times$ 74.0\arcsec & 1.65 mJy~beam$^{-1}$ & (3)
    \end{tabular}}
    \vspace{3mm}
    \tablefoot{The HBA image at a resolution of 145 arcsecond was smoothed from the 60 arcsecond tapered image. (1): \citep{2024A&A...692A..12V} ; (2): \cite{2017MNRAS.469.3872G}; (3): \citep{2011A&A...526A...9B}.}
\end{table*}

We observed the Perseus cluster with the LOFAR LBA system.
We used two observations, one from January 30, 2021, and another from December 29, 2020 (project code LC15\_027, PI: Osinga for both observations).
The observations employed a standard setup with 64 frequency channels per 195.3\,kHz subband and a 1\,s integration time, recording all four polarization products. The subbands covered a frequency range between 10.4 and 57.7\,MHz.
During the observation, the primary calibrator source 3C\,196 was observed simultaneously using a second station beam.
For the observation on December 29, 2020, the total integration time was 8~hr, and for the observation on January 30, 2021, the total integration time was 4~hr.
For this work, we only use the subbands above 30.0~MHz. Subbands at lower frequencies were discarded due to challenging ionospheric conditions (see below for further details). This resulted in an effective bandwidth of 27.6~MHz, corresponding to data between 30.0 and 57.7~MHz. The visibility data were compressed using Dysco \citep{2016A&A...595A..99O}. 

The data reduction started with the removal of the bright sources Cassiopeia~A and Cygnus~A from visibilities using the demixing procedure \citep{2007ITSP...55.4497V} and flagging of radio frequency interference with AOFlagger \citep{2010MNRAS.405..155O,2010ascl.soft10017O}. Next, instrumental corrections (bandpass, polarization offset) were derived from the calibrator 3C\,196 using the method outlined in \cite{2019A&A...622A...5D}. These corrections were transferred from the calibrator source to the target. The flux density scale of 3C\,196 was set to that of \cite{2012MNRAS.423L..30S}. 
Subsequently, the target field was self-calibrated using direction-independent calibration, and this was followed by direction-dependent calibration.

For the direction-independent (self-)calibration we used a starting model derived from the TIFR GMRT Sky Survey alternative data release 1 \citep[TGSS ADR1;][]{2017A&A...598A..78I}. In addition, we included an extra round of radio frequency interference flagging on the Stokes\,V visibility products with AOFlagger. The self-calibration was performed in a circular correlation basis so that differential Faraday rotation by the ionosphere only manifests itself as a phase difference between the right (RR) and left (LL) handed correlations \citep[][]{2011A&A...527A.106S,2011A&A...527A.107S}. 
Both observations were calibrated individually during the direction-independent self-calibration step.
Due to the presence of the bright radio source 3C\,84 at the centre of the field, the direction-independent calibration mostly optimises the image quality for this source, while other (fainter) sources away from 3C\,84 still suffer from direction-dependent calibration artefacts caused by the ionosphere. 
After the direction-independent calibration step, we discarded data where the ionosphere was restrictively active, which was determined by manual inspection of the direction-independent solutions, leaving 2 hours of the observation from January 30, 2021, and 4 hours of the observation from December 29, 2020.

For the direction-dependent calibration, we initially calibrated both observations separately with \verb+facetselfcal+ \footnote{\url{https://github.com/rvweeren/lofar_facet_selfcal}}\citep{2021A&A...651A.115V}, using a strategy tailored specifically to a such a bright source as 3C\,84. Based on the direction-independent image, the field was divided into ten facets, centred on the locations of bright sources in the field of view. The direction-dependent calibration involved multiple iterations of scalarphase calibration, using varying time intervals and smoothness settings for the core and remote stations. After completing the calibration steps, additional flagging was performed on the residual visibility data. These residual data were created by predicting the model (clean component images), corrupting it with the derived direction-dependent calibration solutions, and subtracting the resulting `corrupted sky visibilities' from the uncorrected visibilities.
Following this flagging step, another round of direction-dependent calibration was carried out using the same strategy, including a final flagging pass on the updated residual data.
Subsequently, both observations were calibrated together after applying the final flags, which ensured that only good data were used for the subsequent self-calibration.

\begin{figure*}[!ht]
    \centering
    \includegraphics[width=\textwidth]{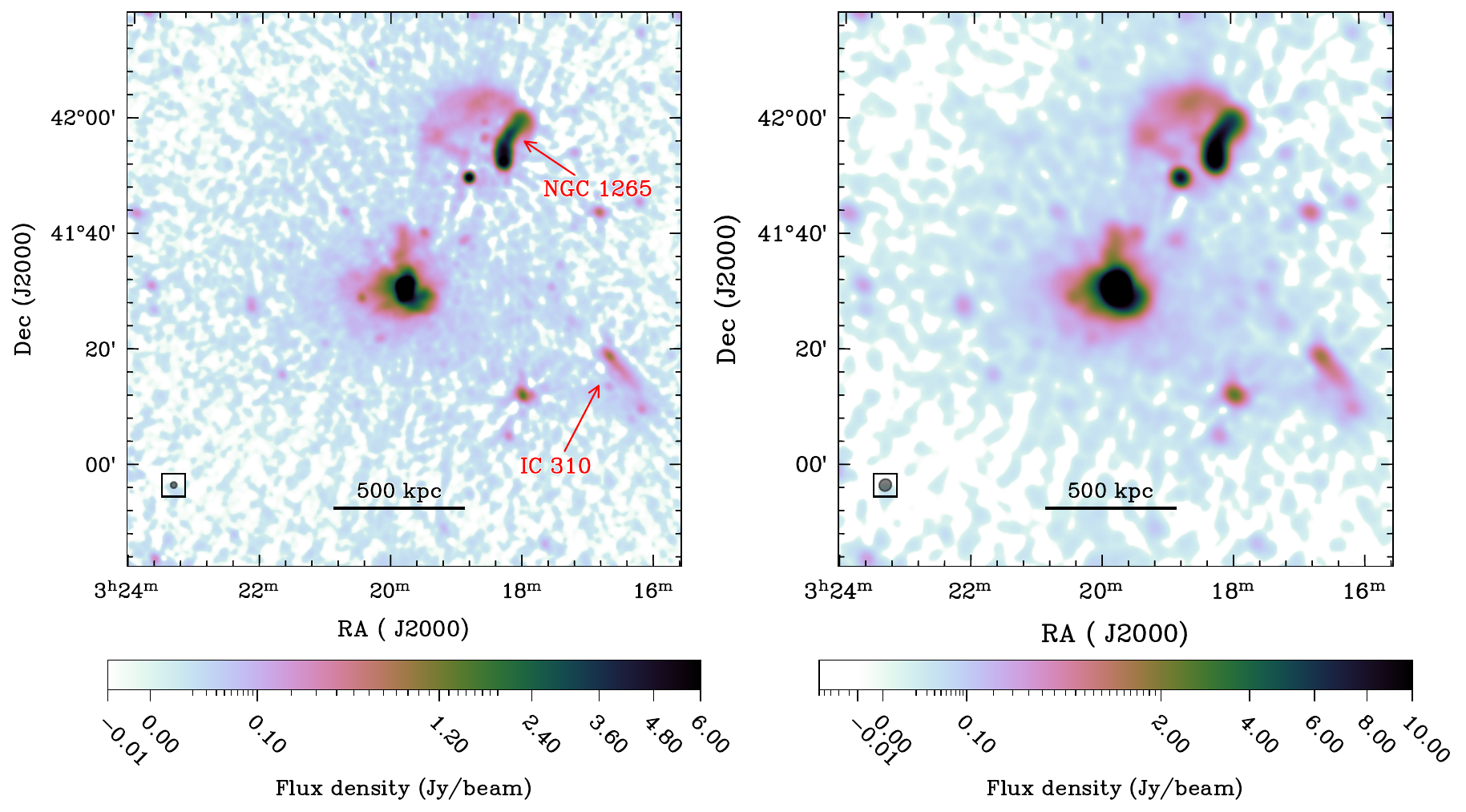}
    \caption{LOFAR LBA image of the Perseus Cluster (30.0--57.7~MHz) tapered to 67\arcsec{} (\textit{left}) and 131\arcsec{} (\textit{right}). The tailed radio galaxies NGC~1265 and IC~310 are marked. For image details, see Table \ref{tab:details}.}
    \label{fig: fullimg_taper}
\end{figure*}

The final images were made using WSClean\footnote{\url{https://gitlab.com/aroffringa/wsclean}, v3.6} \citep{2014MNRAS.444..606O,2017MNRAS.471..301O}. We used a  {multi-scale} bias of 0.75 for the multi-scale clean. An overview of the images and their properties is given in Table \ref{tab:details}.
The LOFAR images we used to compare across frequencies were made with an inner uv cut of $30 \lambda$, which corresponds to the minimal baseline of LOFAR HBA.
The VLA images have larger minimum baselines, but as these images are only used to study smaller-scale emission and not the large-scale diffuse emission in the giant radio halo, the lack of short baselines is not a significant limitation for our VLA images.
The highest resolution LBA image was made with a robust weighting of $-0.5$ and ten facets.
Two additional images at a lower resolution were made using Gaussian uv tapers of 60\arcsec{} and 120\arcsec. This resulted in a beam size of 67.1\arcsec{} and 127.3\arcsec{}, respectively, when taking the average of the minor and major axes (see \autoref{fig: fullimg_taper}). We also include the LOFAR HBA maps from  \cite{2024A&A...692A..12V}. To make spectral index maps, we re-gridded and smoothed these maps to the same resolution as the LBA maps and used the same inner uv cut. In addition to the LOFAR images, we also used the VLA P-band data from \citet[][PI: Hlavacek-Larrondo]{2017MNRAS.469.3872G} and WSRT ($323.9-330.1$~MHz) data from \cite{2011A&A...526A...9B}.

For the radial profiles, we made source subtracted images by first masking out regions with diffuse emission in the model of our original image, then predicting this model, and finally subtracting the model from our data.
The data were re-imaged with a 60\arcsec taper, resulting in a resolution of 67.8\arcsec by 66.4\arcsec and an inner uv cut of 30 $\lambda$ to match the same inner uv cut used for the LOFAR HBA data in \cite{2024A&A...692A..12V}.
The image is presented in Fig. \ref{fig:masked}.

The LBA flux density scale has been shown to be accurate to a level of about 6\% to 10\% \citep{2021A&A...648A.104D,2023A&A...673A.165D}.
As an additional check, we verified the flux density scale of our LOFAR LBA image by comparing the flux density of nine bright sources in the full field of view of the observation with literature values. 
The results are shown in Fig. \ref{fig:flux_check}. The nine sources are marked in Fig. \ref{fig:finder_plot}.
From these results, it appears that in four out of the nine sources, the integrated radio spectra are close to a simple power law shape. 
Four sources display a spectral turnover (3C\,82, 2MASX J03302713+4101418, 4C\,39.11 and NVSS J032136+435923), and one source (NVSS J032707+421925) shows an indication of spectral flattening at low frequencies.
The four sources with a spectral turnover are all located in the periphery of the field of view of the LOFAR image and are at a large distance from the pointing centre. The LBA flux density of the sources are unexpectedly low based on a simple extrapolation from the higher frequency data points. This might have been caused by a poor calibration quality near the edge of the field of view, affecting the flux density of the source.
However, as our science targets are located near the centre of the field of view, we do not expect this to impact our measurements.
NVSS J032707+421925 also shows hints of flattening at slightly higher frequencies, which indicates that this spectrum is regular.
In conclusion, based on this additional check of the flux density scale, we do not find evidence of any global systematic offsets in the pointing centre. Given the challenging calibration and the perturbing effects of 3C\,84, we adopted the more conservative value of 10\% for the flux density scale uncertainty.

\section{Results}
\label{sec:results}

The full-resolution LOFAR LBA image of the cluster core is presented in Fig. \ref{fig: highres_core}. This image still shows minimal signs of scintillation and resulting dynamic range limitations around 3C\,84, which are visible as radial artefact patterns that are mainly present in the high-resolution image. 
Lower-resolution images of the cluster are shown in Fig. \ref{fig: fullimg_taper}, with the 67\arcsec{} resolution image displayed in the left panel and the 131\arcsec{} resolution in the right panel.
Below, we describe the various sources of emission detected in the LOFAR LBA image: the radio mini-halo and the giant radio halo, the ghost cavities near 3C\,84, and the tailed radio galaxies NGC\,1265 and IC\,310.

\subsection{Mini-halo and giant radio halo}

\begin{figure}
    \centering
    \includegraphics[width=\linewidth]{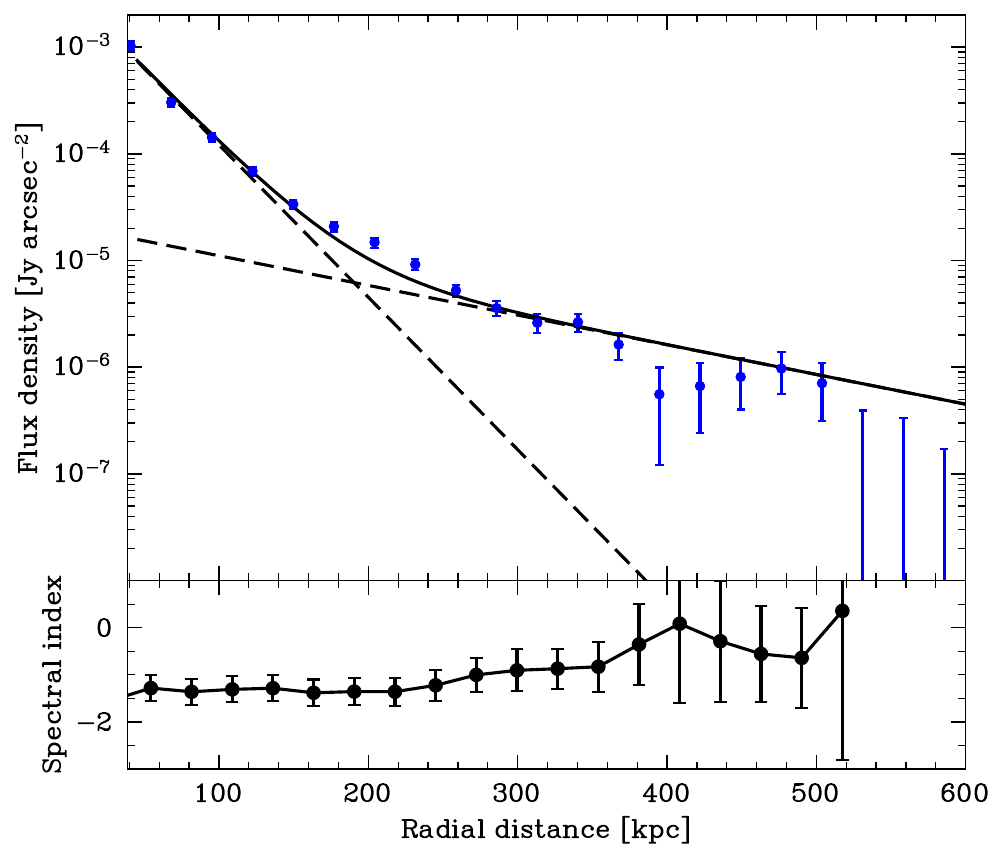}
    \caption{Radial surface brightness profile of the diffuse radio emission in the Perseus cluster. The dashed lines represent two exponential profiles that add up to the continuous line (Eq.~\ref{eq:profile}). The model was fitted using the extracted radial profile of the Perseus image, with point sources subtracted and residual sources masked out, starting from 43~kpc from the centre to 650~kpc. We note that the error bars do not include the 10\% uncertainty on the adopted flux density scale, as this uncertainty is identical between all data points. In the bottom panel, the spectral index between LOFAR LBA (43~MHz) and LOFAR HBA (144~MHz) is shown with error bars (which do not include the 10\% uncertainty). We note that at larger radial distances from 3C\,84, the error on the spectral index drastically increases.}
    \label{fig:lba_profile}
\end{figure}

The most prominent source of diffuse radio emission in the centre of the cluster is the well-studied radio mini-halo \citep{1975A&A....45..223M,1983SvAL....9..305S,2017MNRAS.469.3872G}.
In our LOFAR LBA images, the radio mini-halo is clearly detected, and it displays an overall similar morphology as in \cite{2024A&A...692A..12V}, with a pronounced region of diffuse emission surrounding the brightest cluster galaxy NGC\,1275 and the associated radio source 3C\,84.
Some of the structures that are detected in \cite{2024A&A...692A..12V} and in \cite{2017MNRAS.469.3872G} are also visible in the LOFAR LBA map, namely, the \enquote{Northern extension} and \enquote{Eastern spur S}, which are marked in Fig. \ref{fig: highres_core}.
To the east of the core of the cluster, several filament-like structures can be seen, similar to \cite{2024A&A...692A..12V}.
However, these structures appear affected by calibration artefacts from 3C\,84, which limit the dynamic range of the image.
Most of the filamentary substructures and sharp surface brightness edges in this emission reported in \cite{2024A&A...692A..12V} are not detected in the LBA image. 
This can be explained by the lower sensitivity and more challenging ionospheric conditions of the LBA observations.

On scales beyond the mini-halo, we noted the presence of faint diffuse emission in our 67\arcsec{} and 131\arcsec{} resolution LBA images (see Fig. \ref{fig: fullimg_taper}).
This emission was identified as a giant radio halo by \cite{2024A&A...692A..12V}.
The giant radio halo becomes dominant over the mini-halo at a distance of $0.2R_{500}\approx \SI{260}{\kilo\parsec}$ \citep{2024A&A...692A..12V}, and it has not been detected in previous VLA, GMRT, and WSRT observations, possibly due to the limited sensitivity of these telescopes to large-scale diffuse emission in combination with a steep radio spectrum.

To quantify the two different radio halo components and their radial brightness distribution, we fitted a two-component exponential function to the radio emission in the 68\arcsec{} LOFAR LBA map, with point sources subtracted (see Fig. \ref{fig:masked}).
The LOFAR LBA image is worse in terms of residual ionospheric effects compared to the LOFAR HBA image.
Therefore, we fixed the radius of the inner (MH) component and the outer (GRH) component to the values obtained in \cite{2024A&A...692A..12V}.
The results are presented in Fig. \ref{fig:lba_profile}, where we fitted the following double exponential profile to the data:
\begin{align} I_{\nu,tot}(r) = I_{\nu,0,in}\text{exp}\left(-\frac{r}{r_{e,in}}\right) + I_{\nu,0,out} \text{exp}\left(-\frac{r}{r_{e,out}}\right).
\label{eq:profile}
\end{align}

Here, $I_{\nu,0,in}$ and $r_{e,in}$ respectively refer to the central brightness and e-folding radius of the inner component (likely corresponding to the minihalo), whereas $I_{\nu,0,out}$ and $r_{e,out}$ respectively refer to the central brightness and e-folding radius of the outer component (likely corresponding to the giant radio halo).
We set $r_{e,in}=30.4\pm0.2 \,\text{kpc}$ and $r_{e,out} = 156\pm 2 \,\text{kpc}$.
After fitting this profile, we obtained the following values:
\begin{table}[H]
    \centering
    \begin{tabular}{l}
    $I_{\nu,0,in} = \SI{3.0\pm 0.31}{\milli\jansky\per\asec\squared}$ \\
    $I_{\nu,0,out} = \SI{35 \pm 3.6}{\micro\jansky\per\asec\squared}$ 
    
    \end{tabular}
\end{table}

The two-component model provides a good description of the observed radial profile.
To compare the total power of the mini-halo and the giant radio halo, we integrated the intensity of the inner component to a radius of 150~kpc and the intensity of the outer component to a radius of 750~kpc, corresponding to $5\times r_e$, using the value of $r_e$ as reported in \citep{2024A&A...692A..12V}.
The values for the flux density and the uncertainties are reported in Table \ref{tab: fluxdens}.

\begin{table*}[tbp]
    \caption{Integrated flux densities of the diffuse components in the Perseus cluster.}
    \label{tab: fluxdens}
    \centering
    \begin{tabular}{c|c|c|c}
         & LBA (43~MHz) & HBA (144~MHz)  & $\alpha$\\ \hline
         Mini-halo & $126.7\pm 4.2 \pm 13.4$ Jy & $25.3\pm 0.67 \pm 1.41$ Jy & $-1.34 \pm 0.10$\\
         GRH & $21.4 \pm 1.0 \pm 2.4$ Jy & $ 6.34\pm 0.26 \pm 0.41$ Jy & $-1.01 \pm 0.11$
    \end{tabular}
    \tablefoot{Integrated flux densities are determined in this work and \cite{2024A&A...692A..12V}. Uncertainties are twofold: The first uncertainty represents the one reported by the Chi-squared fit (Eq.~\ref{eq:profile}) and is based on the r.m.s. map noise, and the second refers to the systematic 10\% flux density scale uncertainty. The error on the spectral indices in this table includes both uncertainties. The uncertainties on the spectral indices without the 10\% flux density uncertainties are $\pm 0.035$ for the mini-halo and $\pm 0.051$ for the giant radio halo.}
\end{table*}

\begin{figure}
    \centering
    \includegraphics[width=0.75\linewidth,trim={0cm 0cm 12cm 0cm},clip]{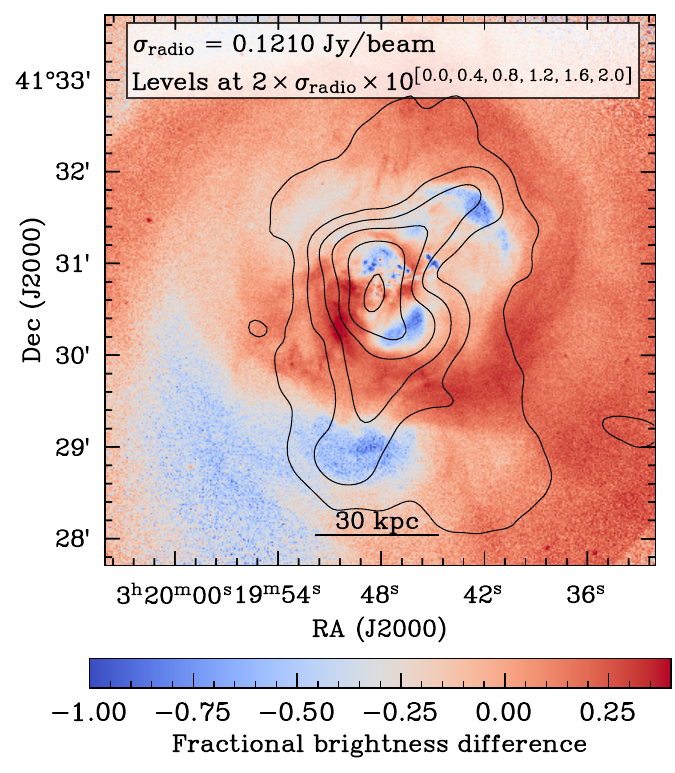}
    \caption{Comparison between radio emission detected with LOFAR LBA (native resolution) and X-ray emission (with a radially smooth profile subtracted). Radio contours are drawn at the levels labelled in the image.}
    \label{fig:xrayoverlay}
\end{figure}

The inner mini-halo component has an integrated flux density of $126.7\pm 13.4$ Jy. For the outer giant radio halo component, we found an integrated flux density of $21.4\pm 2.4$ Jy.
Combining the LBA measured flux densities with the published HBA profile fitting results, the mini-halo in the Perseus cluster has a spectral index of $\alpha_{MH} = -1.34 \pm 0.11$, which is slightly steeper compared to earlier measurements of the spectral index of the mini-halo ($-1.10\pm0.05$ between 143~MHz and 352~MHz, \citealt{2024A&A...692A..12V}; and $-1.21\pm 0.05$ between 325 MHz and 1.4 GHz, \citealt{1993PhDT.......392S,2014ApJ...781....9G}).
The detection of the giant radio halo in both the LBA and HBA observations of the Perseus cluster allowed us to determine the spectral index of the giant radio halo for the first time.
The spectral index of the giant radio halo between 43~MHz and 144~MHz is $\alpha_{GRH} = -1.01 \pm 0.11$.
The errors on these spectral indices include the 10\% flux density scale uncertainty, as discussed in Sect.~\ref{sec:methods}.
This result shows that it is unlikely that the giant radio halo in the Perseus cluster is an ultra-steep spectrum radio halo.
Regarding the difference in the spectral index between the mini-halo and giant radio halo, we found $\Delta_\alpha = \alpha_{GRH} - \alpha_{MH} = 0.33 \pm 0.062$.
In this case, the uncertainty on the difference in the spectral indices does not use the 10\% flux density scale uncertainty, as we are comparing the spectral indices of two spatially different objects (and the flux density uncertainty is correlated one-to-one between the mini-halo and the giant radio halo).
However, it is important to note that due to ionospheric distortions, the diffuse emission could be contaminated by imperfect point source subtraction.
Observations under more favourable ionospheric conditions will likely allow us to investigate the spectral properties of the giant radio halo in more detail.

\begin{figure}
    \centering
    \includegraphics[width=0.85\linewidth]{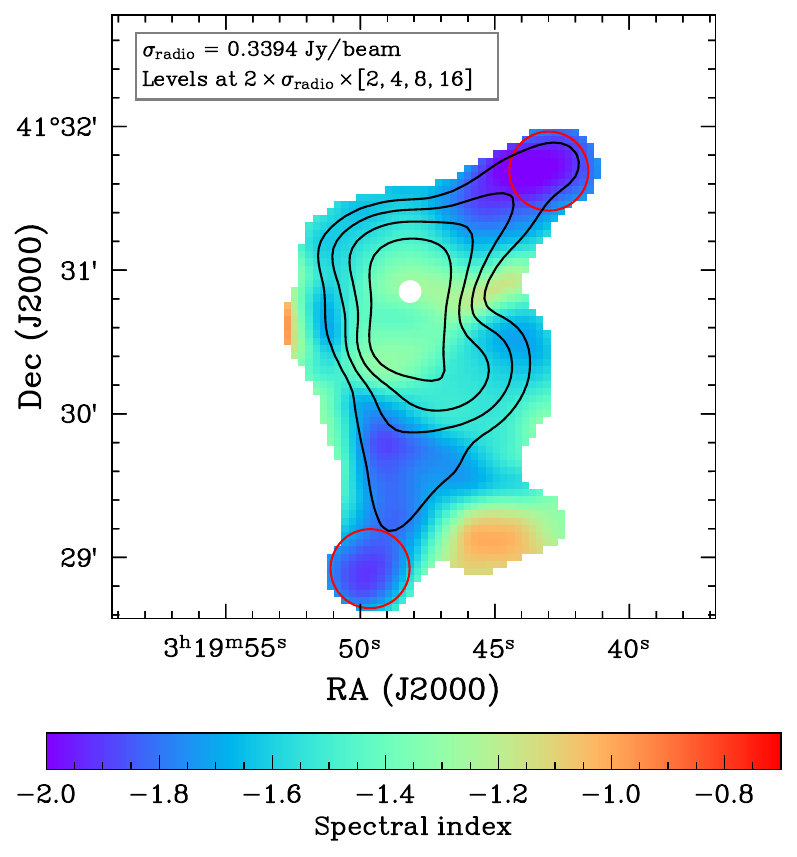}
    \caption{Spectral index map of the minihalo of the Perseus cluster between 44 and 144~MHz. The HBA map has been smoothed to the resolution of the native LBA map. The associated spectral index uncertainty map is displayed in Fig. \ref{fig:spix_err}. We note that 3C\,84 is located at the centre, and it has been blanked out for clarity. The red circles refer to the ghost cavities whose spectra are shown in Fig. \ref{fig:cavity spectrum}. The noise in this image is higher than the noise reported in Table \ref{tab:details} due to the proximity of 3C\,84.}
    \label{fig:zoominspix}
\end{figure}

The spectral index of the giant halo and mini-halo can also be calculated via an alternative method without fitting the double exponential profile. For this,
we computed the spectral index in two annuli: one with an inner cut of 58~kpc and an outer cut of 90~kpc (corresponding to the region where the mini-halo dominates the diffuse radio emission) and the other with an inner cut of 200~kpc and outer cut of 350~kpc (corresponding to the giant radio halo).
This calculation gave us an integrated spectral index of $\alpha=-1.25 \pm 0.12$ for the mini-halo and $\alpha = -1.28 \pm 0.15$ for the giant radio halo. 
For the radio mini-halo, this measurement of the spectral index is consistent with the measurement based on fitting two profiles. For the giant radio halo, the measurement differs ($\Delta = 0.27 \pm 0.12$) in a statistically significant way.
This discrepancy is probably caused by imperfect source subtraction or interference due to the Milky Way.

\subsection{Ghost cavities}

\begin{figure}
    \centering
    \includegraphics[width=0.9\linewidth]{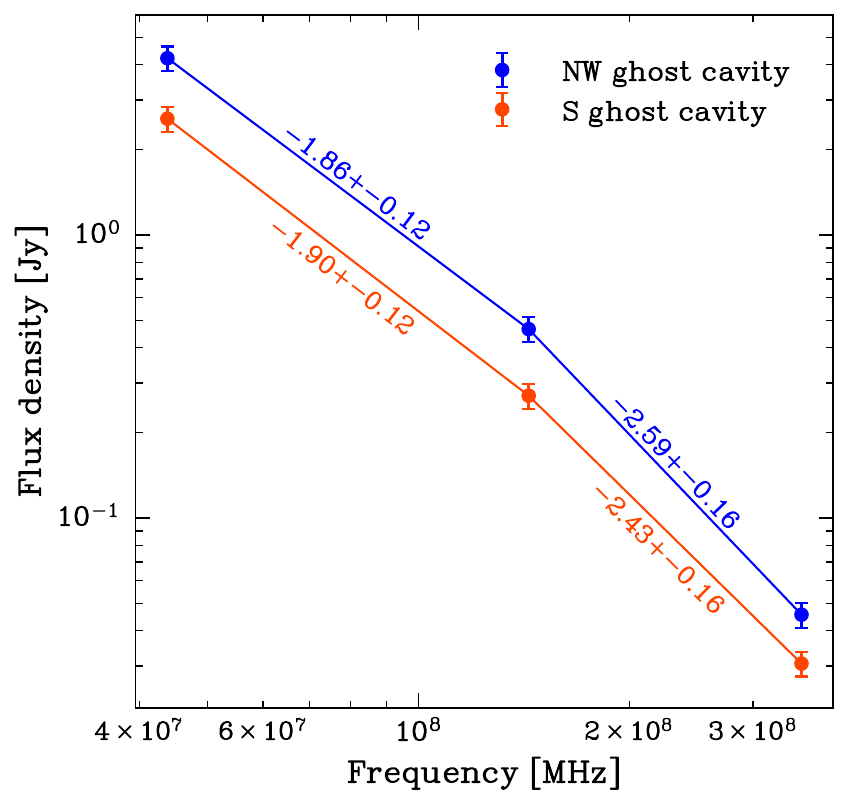}
    \caption{Radio spectrum of the ghost cavities within the regions marked by circles in Fig. \ref{fig:zoominspix}. Piecewise spectral indices are given between the LOFAR LBA, HBA, and VLA P-band observations.}
    \label{fig:cavity spectrum}
\end{figure}

X-ray observations \citep[e.g.][]{1993MNRAS.264L..25B,2000A&A...356..788C,1981ApJ...248...47F,2003MNRAS.344L..43F,2006MNRAS.366..417F,2011MNRAS.418.2154F} have previously revealed two cavities in the X-ray emission in the centre of the Perseus cluster without any corresponding gigahertz radio emission \citep[i.e. so-called ghost cavities;][]{2002IAUS..199..189B,2007ApJS..172..686K}.
In Fig. \ref{fig:xrayoverlay}, we present a fractional difference X-ray map from Chandra with LOFAR LBA contours (at native resolution) similar to \cite{2024A&A...692A..12V} and X-ray data processed by \cite{2011MNRAS.418.2154F}.
The fractional difference X-ray map shows a total of four cavities: two cavities located near 3C\,84 and two ghost cavities located further from the central AGN.
Our LBA image reveals that the radio emission in the core has spurs in the direction of the X-ray ghost cavities and that there is emission filling the cavities, similar to \citep{2024A&A...692A..12V}. To determine the spectral properties of the ghost cavities, we measured the flux densities in the ghost cavities in a circular region centred on the cavity, marked with a red circle in Fig. \ref{fig:zoominspix}.
Within the circle, the northwest cavity has an integrated flux density of $\SI{4.2\pm0.42} {\jansky}$, while the southern region has an integrated flux density of $\SI{2.6\pm0.26} {\jansky}$.
In addition, when comparing the left image of Fig. \ref{fig:xrayoverlay} with the spectral index map of the inner region of the cluster presented in Fig. \ref{fig:zoominspix}, we observed that the radio emission in the X-ray cavities has a steep spectrum with $\alpha=-1.86\pm 0.12$ in the northwest cavity and $\alpha=-1.90 \pm 0.12$ in the southern cavity, as measured by integrating the flux densities within the red circles in Fig. \ref{fig:zoominspix}.
Using the VLA P-band (central frequency of 352~MHz) image as presented in \cite{2017MNRAS.469.3872G} and the LOFAR HBA image, we found $\alpha=-2.59 \pm0.16$ for the northwest ghost cavity and $\alpha=-2.43\pm 0.16$ for the southern ghost cavity.
For both ghost cavities, this means that the spectrum steepens significantly ($\Delta_{\alpha,\text{NW}} = 0.73 \pm 0.2$; $\Delta_{\alpha,\text{S}} = 0.53 \pm 0.2$) between 352~MHz and 43~MHz, as can be seen in Fig. \ref{fig:cavity spectrum}.

\begin{figure*}
    \centering
    \includegraphics[width=0.48\linewidth]{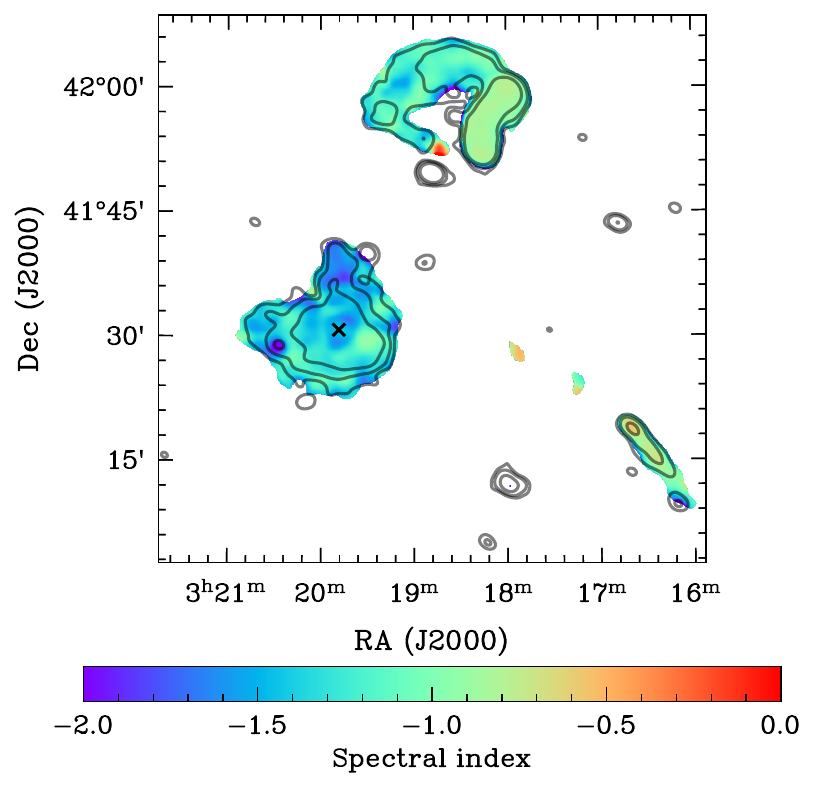}
    \includegraphics[width=0.48\linewidth]{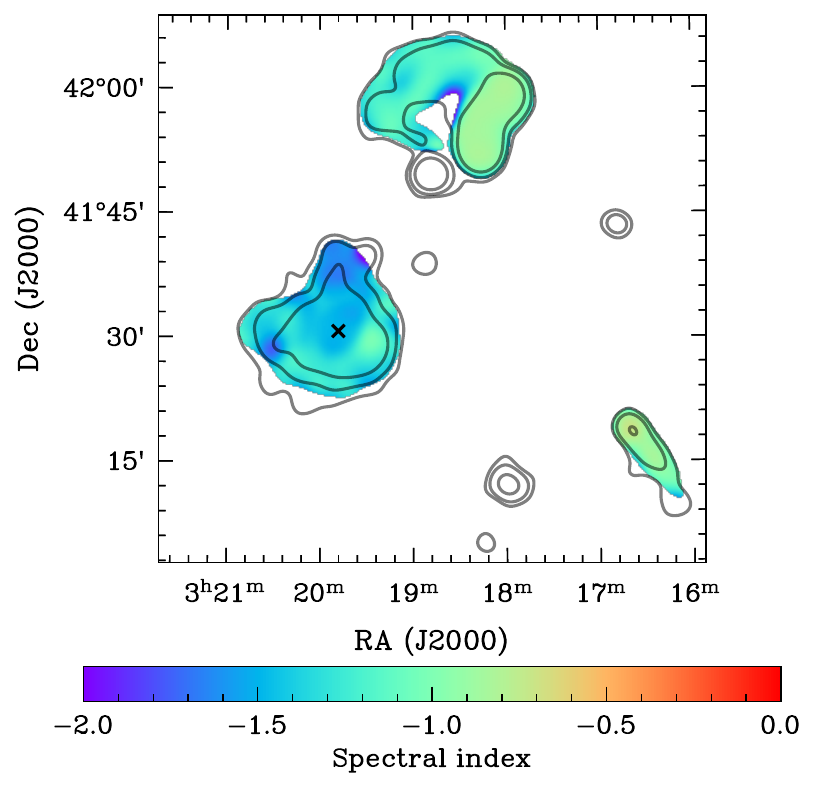}
    \caption{LBA-HBA spectral index maps of the Perseus cluster, including the tailed radio galaxies NGC~1265 and IC~310, between LOFAR LBA and LOFAR HBA frequencies. The images have been smoothed and re-gridded to the same resolution.
    \textit{Left}: Spectral index map at 67\arcsec{} resolution.
    The black contours are from the LBA image and drawn at levels of $\sigma_{LBA} \times \sqrt{[128,512,4096]}; \sigma_{LBA}=\SI{0.0127}{\jansky\per\beam}$.
    \textit{Right}: Spectral index map at 131\arcsec{} resolution. 
    The black contours are from the LBA image and drawn at levels of $\sigma_{LBA} \times \sqrt{[128,512,4096]}; \sigma_{LBA}=\SI{0.0266}{\jansky\per\beam}$.
    The black cross marks the centre of the cluster.
    For more detail about these images, see Table \ref{tab:details}. The corresponding spectral index uncertainty maps are shown in Fig. \ref{fig:spixerr}.}
    \label{fig:spix_map_large}
\end{figure*}

\subsection{NGC 1265 and IC 310}

Apart from the mini-halo, giant halo, and 3C~84, the Perseus cluster hosts two other extended radio sources that are prominently detected in our LOFAR LBA images: NGC~1265 and IC~310.
These sources are annotated in Fig. \ref{fig: fullimg_taper}.
In addition, a spectral index map of the Perseus cluster including NGC~1265 and IC~310 is shown in Fig. \ref{fig:spix_map_large}.

\begin{figure*}
    \centering
    \includegraphics[width=0.9\linewidth]{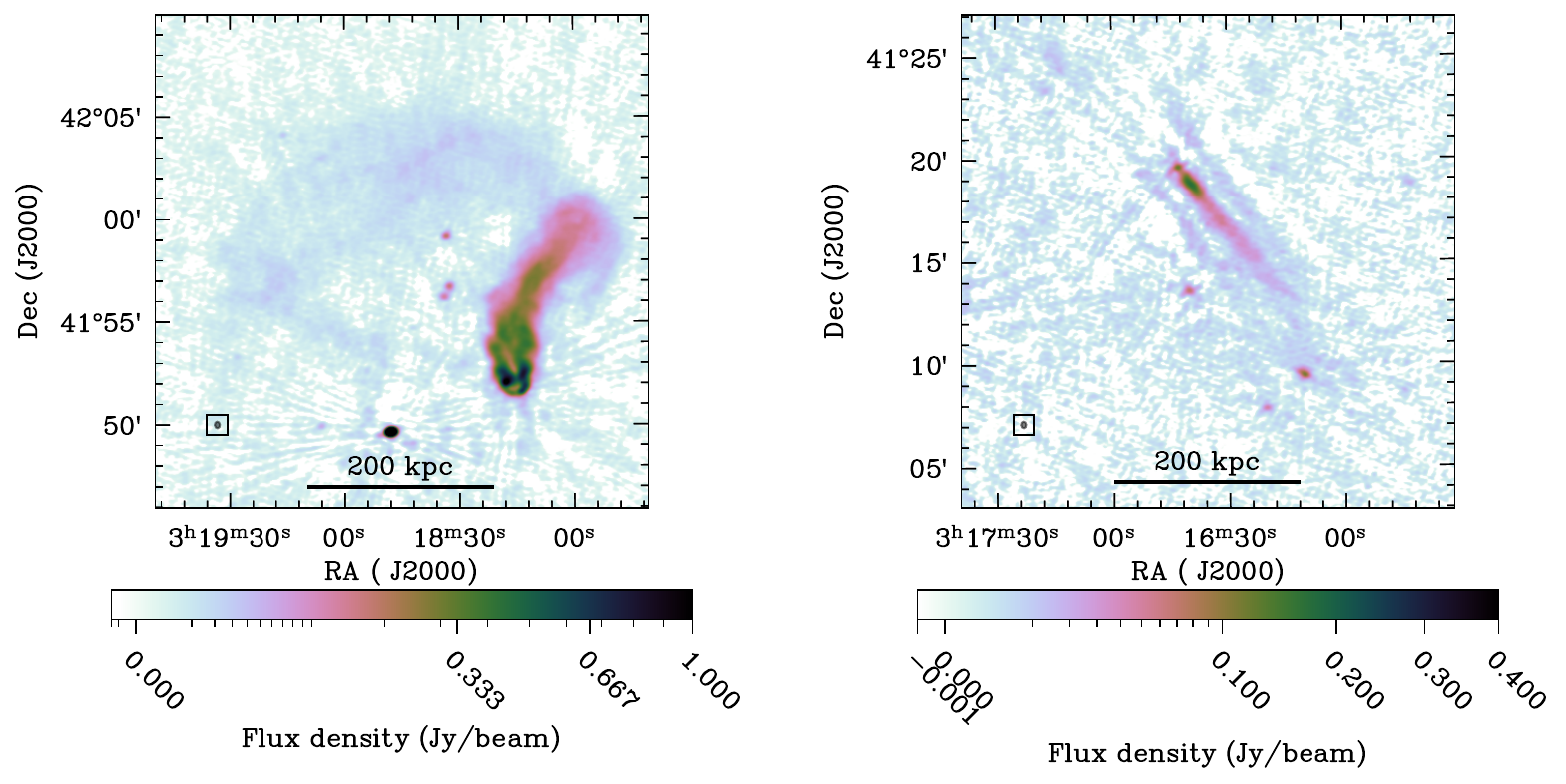}
    \caption{Zoomed-in full resolution LBA images of NGC 1265 (\textit{left}) and IC~310 (\textit{right}).}
    \label{fig:agn}
\end{figure*}

\begin{figure*}
    \centering
    \includegraphics[width=0.45\textwidth]{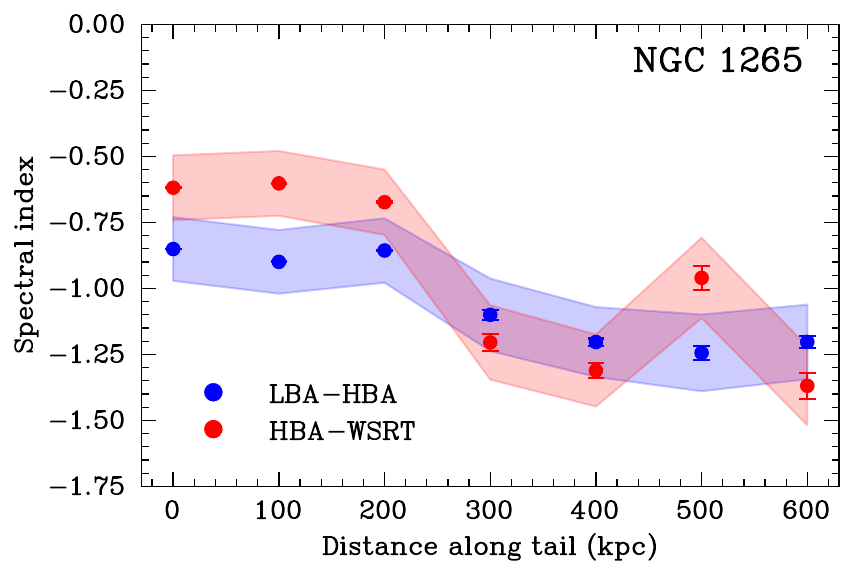}
    \includegraphics[width=0.45\textwidth]{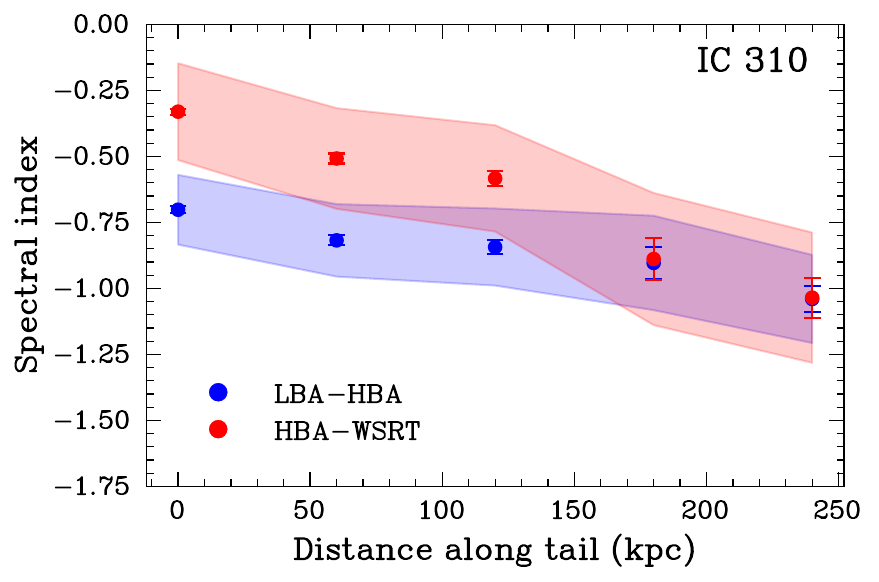}
    \caption{Spectral indices along the tails of NGC~1265 and IC~310. The flux densities are measured in circular regions with radii of 1.8\arcmin and 1\arcmin at equal spacings of 100~kpc and 60~kpc for NGC~1265 and IC~310, respectively. These regions are marked with circles in the cutout images presented in Figs. \ref{fig: ngcthrice} and \ref{fig: icthrice}. The error bars indicate the spectral index uncertainty when taking only the r.m.s. map noise into account (relevant for comparing the evolution of the spectral index along a spatial dimension), while the shaded regions include the 10\% flux density uncertainty (relevant for comparing the spectral index between different frequencies).}
    \label{fig: profiles}
\end{figure*}

The source NGC~1265 is a narrow-angle tail radio galaxy with a peculiar morphology.
A full-resolution image of this tail is shown in the left panel of Fig. \ref{fig:agn}.
The source has two jets extending in the northern direction, and they merge to become one tail moving in the northwestern direction.
Subsequently, the tail bends to the east and eventually south, and it then becomes much fainter. In Fig. \ref{fig: profiles} (left panel), we present the spectral index of NGC~1265 along the tail.
Spectral indices are measured in circles of 1.8\arcmin (40~kpc) spaced 4.5\arcmin (100~kpc) apart on the tail (based on the cluster redshift).
Flux densities were measured from the 43~MHz LBA, 144~MHz HBA, and 352~MHz WSRT images.

The spectral index between LOFAR LBA (43~MHz) and LOFAR HBA (144~MHz) stays roughly constant (around $-0.8$) up to about 200~kpc, where it subsequently becomes significantly steeper, with $\alpha\approx-1.2$.
This transition corresponds to the break between the initial northwestern tail and the east and southern extension, as can be seen in Fig. \ref{fig:spix_map_large}.
We found a similar evolution of the spectral index between HBA and WSRT frequencies.
We did not detect any significant spectral curvature along the tail.
Interestingly, NGC\,1265 appears to host an inverted spectrum close to the head of the radio galaxy, although this is only marginally significant.

The source IC~310 is a narrow-angle tail radio galaxy in the Perseus cluster \citep{2001ApJS..134..355M,2016AJ....152...50T} that is likely radially infalling \citep{2020MNRAS.499.5791G}.
It has been suggested that IC~310 is a low-luminosity FRI galaxy whose jets are aligned towards our line of sight \citep{1999ApJ...516..145R,2020MNRAS.499.5791G}.
We present a full-resolution image of IC~310 in the right panel of Fig. \ref{fig:agn}.
Spectral indices were measured in circles of 1.5\arcmin (33~kpc) spaced 2.7\arcmin (60~kpc) apart on the tail.
IC~310 displays a straight tail with a sharp bend towards the centre of the cluster at around 200~kpc \citep{2020MNRAS.499.5791G}.

Similar to NGC\,1265, IC\,310 appears to host an inverted spectrum near the tip. This could also be the head of the radio galaxy.
\cite{2024A&A...692A..12V} has shown that the emission of IC~310 continues after this bend until it is fully blended with the diffuse emission in the cluster centre.

In our work, we detected IC~310 as a 12\arcmin{} elongated source (corresponding to 250~kpc).
The spectral index profile (Fig. \ref{fig: profiles}, right panel) shows that the spectral index steepens along the tail, from around -0.75 at the beginning of the tail to -1.25 at 240~kpc.
The spectral index steepens linearly along the source (Fig. \ref{fig:spix_map_large}).
We observed some evidence of the earlier mentioned bend in IC~310, but it is significantly less pronounced compared to the HBA image.
Similar to NGC\,1265, IC~310 appears to host an inverted spectrum near the head of the radio galaxy.

\section{Discussion}
\label{sec:discussion}
\subsection{Large-scale diffuse emission}
The classical picture of cluster-scale diffuse emission associates radio mini-halos with cool core (relaxed) clusters and giant radio halos to disturbed systems \citep[e.g.][]{2012A&ARv..20...54F,2010ApJ...721L..82C}.
However, observations have revealed that some clusters that appear to be in an intermediate stage between a cool core and a non-cool core cluster can host both a mini-halo and large-scale extended emission \citep{2014MNRAS.444L..44B,2017A&A...603A.125V,2021MNRAS.508.3995B,2022A&A...657A..56K,2023A&A...678A.133B,2024A&A...683A.132L}.
In addition, \cite{2024A&A...686A..82B} have found indications that sloshing features in the cluster core, typically associated with radio mini-halos, are related to the presence of larger-scale emission beyond the scales of mini-halos.

A weak lensing analysis of the Perseus cluster has shown that the cluster is undergoing a major merger (mass ratio 3:1) with a large impact parameter \citep{2025NatAs...9..925H}.
This indicates that the Perseus cluster is in an intermediate phase. Due to the merger, cosmic rays in the cluster periphery could experience turbulent re-acceleration, giving rise to a giant radio halo commonly associated with disturbed systems.
The centre of the Perseus cluster, however, remains unperturbed for the time being, and here gas sloshing motions are re-accelerating cosmic rays to form the mini-halo (associated with non-disturbed systems). 
This has also been seen in similar cases where despite evidence of a merger, the cool-core of the galaxy cluster remains intact \citep{2018ApJ...868..121D,2010A&A...510A..83R}.
Alternatively, the mini-halo could trace cosmic ray electrons from hadronic interactions.

We measured a low-frequency spectral index of $\alpha=-1.34 \pm 0.10$ for the radio mini-halo, consistent with previous measurements at higher frequencies within the uncertainties \citep[][]{1993PhDT.......392S,2021ApJ...911...56G,2024A&A...692A..12V}. For the giant radio halo, the measured spectral index of $\alpha=-1.01 \pm 0.12$ is not in line with the expectation from turbulent re-acceleration models that giant radio halos in such intermediate phase clusters have a relatively steep spectral index \citep{2017A&A...603A.125V,2018MNRAS.478.2234S,2021MNRAS.508.3995B,2023A&A...678A.133B,2024A&A...686A..44R}.
We note, however, that our measurements may suffer from interference from the Milky Way, as the Perseus cluster has a relatively low galactic latitude ($l = -13{\degree}$), and  the spectral index of the large-scale diffuse emission from the Milky Way is ultra steep  \citep[$\alpha\approx -2.5$;][]{2017MNRAS.464.3486Z,2025arXiv251208522I}.
In addition, it is possible that the spectrum of the giant radio halo flattens at a lower frequency, and we are observing this flatter regime \citep[e.g. as is the case with the Coma cluster][]{2003A&A...397...53T}. 
Sensitive observations at higher frequencies, such as with the MeerKAT telescope, could potentially detect the steeper part of the integrated spectrum of the giant radio halo.
Based on the LOFAR HBA data, and assuming a spectral index of -1.5, we expect the giant radio halo to have a radio intensity at 800~MHz (UHF) and at $2r_e$ of $\SI{12.4}{\micro\jansky\per\beam}$.

This work has been unable to probe most filaments and edges in the radio-mini halo detected in previous studies.  \cite{2017MNRAS.469.3872G,2024A&A...692A..12V} has shown that there are several filaments and edges present in the mini-halo of the Perseus cluster. %
The lack of a detection in our LBA image can be attributed to the relatively low signal-to-noise ratio and resolution achieved for the radio mini-halo.
These properties are due to the limited image fidelity and challenging ionospheric conditions. To better study the mini-halo morphology, the giant radio halo, and their spectral index distributions, significantly more integration time is required under better ionospheric conditions and ideally with an improved resolution such that point sources can be better subtracted from the data.
However, the `Northern extension' (as marked in Fig. \ref{fig: highres_core}) is clearly visible in LBA, which suggests that this filament has a relatively steep spectrum (as can be seen in Fig. \ref{fig:spix_map_large}). 

Incorporating the European baselines of LOFAR would enable more accurate subtraction of 3C\,84 and thereby improve the dynamic range in the cluster core, as demonstrated in \cite{2024A&A...692A..12V}. 
For LBA, narrowband observations have already been taken of the Perseus cluster \citep{2025A&A...704A..65B}, but calibration of international baselines is currently still restrictively difficult \citep[e.g.][]{2016MNRAS.461.2676M,2022A&A...658A...9G}.
Additionally, achieving more precise spectral index maps will require a broader frequency baseline. However, this poses challenges due to the limited sensitivity achievable with LOFAR’s decameter-band observations \citep{2024NatAs...8..786G} or higher-frequency P-band VLA and uGMRT observations. The VLA offers limited P-band sensitivity to steep-spectrum emission, and the giant radio halo was not detected by \cite{2017MNRAS.469.3872G}. Observations from the upgraded Giant Metrewave Radio Telescope (uGMRT) Band 3 face significant challenges due to the high dynamic range needed for 3C\,84 and the limited number of short baselines available to capture the very extended emission of the giant radio halo.

\subsection{Ghost cavities}
Through our LBA images, we confirm the presence of steep spectrum emission in the ghost cavities near 3C\,84. It is expected that these ghost cavities contain aged AGN plasma originating from jets of 3C\,84 that initially inflated two cavities that subsequently drifted buoyantly to their current locations \citep[e.g.][]{
2001ApJ...554..261C,2002Natur.418..301B,2003ApJ...592..839B,2008ApJ...685..128M}.

The above scenario would imply that the emission in the ghost cavities is steeper than the radio emission associated with the inner pair of X-ray cavities. Indeed, HBA observations \citep{2024A&A...692A..12V} have shown this to be the case. In our work, we have established that the radio spectral index is lower between 144 and 352~MHz compared to 44 and 144~MHz. This spectral steepening towards higher frequencies is expected if the plasma in the ghost cavities has undergone spectral aging. Future LBA observations with European baselines would be important to carry out detailed resolved studies of the ghost and cavity pairs of 3C\,84.

\section{Conclusions}
\label{sec:conclusion}

This paper presents the deepest and highest resolution observations of the Perseus galaxy cluster below 100~MHz to date. Using LOFAR LBA observations spanning 30.0--57.7~MHz with a resolution of $19.2\arcsec\times 15.0\arcsec$, we investigated the diffuse emission in the core of the Perseus cluster as well as the two tailed radio galaxies, NGC~1265 and IC~310. Additionally, we produced lower angular resolution images to analyse the radio emission on larger angular scales. Below we list the main conclusions of our work:

\begin{itemize}
    \item We detected both the radio mini-halo, which has been studied extensively at higher frequencies, and the recently discovered giant radio halo.
    Combining our LBA images with previously published LOFAR HBA observations, we measured an integrated spectral index of $-1.34\pm 0.10$ between 144 and 44\,MHz for the radio mini-halo. For the giant radio halo, we found a spectral index of $-1.01\pm 0.12$ when measuring the spectral index with a radial fit and a spectral index of $-1.28\pm 0.15$ when measuring the spectral index in an annulus between 200 and 350 kpc. 
    The discrepancy is probably caused by ionospheric distortions, which can cause complications for source subtraction.
    \item We detected steep spectrum radio emission in the two X-ray ghost cavities in the core of the Perseus cluster. Between 44 and 144\,MHz, we measured spectral indices of $-1.86\pm0.12$ and $-1.90\pm 0.12$ for the northwestern and southern ghost cavity, respectively.
    Combining LBA, HBA, and VLA P-band observations, we characterised the radio spectral shape and found evidence of the steepening of the radio spectrum towards higher frequencies.
    
    \item In the LBA observations, we detected two tailed radio galaxies in the outskirts of the Perseus cluster, NGC\,1265 and IC\,310. Including HBA observations, both sources show an indication of spectral steepening along the tails. After bending, the second fainter part of the NGC\,1265 tail shows a relatively constant steep spectral index of $\approx -1.2$.
    
\end{itemize}

This work emphasises the importance of sub-100-MHz LBA observations of galaxy clusters with LOFAR, as they complement LOFAR HBA observations for spectral studies. Future LBA observations of the Perseus cluster with improved uv coverage and longer integration times -- beyond the currently used 6 hours -- will be crucial for detecting substructures in the mini-halo and reducing the uncertainty on the spectral index of the giant radio halo and will thus provide better constraints on its origin. Similarly, a detection at frequencies higher than the HBA band will be important. However, this will be challenging with current telescopes. For instance, the VLA offers limited P-band sensitivity to steep-spectrum emission, while uGMRT also faces significant challenges due to the very high dynamic range requirements needed for 3C\,84 and the limited number of short baselines of uGMRT necessary to capture the very extended emission of the giant halo. Finally, extending the frequency coverage to the decameter band would enable better characterisation of the spectral shape of the radio emission.

\begin{acknowledgements}
CG and RJvW acknowledge support from the ERC Starting Grant ClusterWeb 804208.
CG, FdG and GDG acknowledge the support of the ERC Consolidator Grant ULU 101086378.
MLGM acknowledges financial support from NSERC via the Discovery grant program and the Canada Research Chair program. The Dunlap Institute is funded through an endowment established by the David Dunlap family and the University of Toronto. 
This paper is based (in part) on data obtained with the International LOFAR Telescope (ILT) under project code \verb+COM_LBA_SPARSE+. LOFAR \citep{2013A&A...556A...2V} is the Low Frequency Array designed and constructed by ASTRON. It has observing, data processing, and data storage facilities in several countries, that are owned by various parties (each with their own funding sources), and that are collectively operated by the ILT foundation under a joint scientific policy. The ILT resources have benefitted from the following recent major funding sources: CNRS-INSU, Observatoire de Paris and Universit\'e d'Orl\'eans, France; BMBF, MIWF-NRW, MPG, Germany; Science Foundation Ireland (SFI), Department of Business, Enterprise and Innovation (DBEI), Ireland; NWO, The Netherlands; The Science and Technology Facilities Council, UK.
This research made use of the LOFAR-IT computing infrastructure supported and operated by INAF, including the resources within the PLEIADI special ``LOFAR'' project by USC-C of INAF, and by the Physics Dept. of Turin University (under the agreement with Consorzio Interuniversitario per la Fisica Spaziale) at the C3S Supercomputing Centre, Italy. 
This work made use of Astropy:\footnote{\url{http://www.astropy.org}} a community-developed core Python package and an ecosystem of tools and resources for astronomy \citep{2013A&A...558A..33A,2018AJ....156..123A,2022ApJ...935..167A}
\end{acknowledgements}

\bibliographystyle{aa}
\bibliography{processed_main}

\begin{thebibliography}{104}
\expandafter\ifx\csname natexlab\endcsname\relax\def\natexlab#1{#1}\fi

\bibitem[{{Acciari} {et~al.}(2009){Acciari}, {Aliu}, {Arlen}, {Aune},
  {Bautista}, {Beilicke}, {Benbow}, {Boltuch}, {Bradbury}, {Buckley}, {Bugaev},
  {Byrum}, {Cannon}, {Celik}, {Cesarini}, {Ciupik}, {Cogan}, {Cui},
  {Dickherber}, {Duke}, {Fegan}, {Finley}, {Fortin}, {Fortson}, {Furniss},
  {Galante}, {Gall}, {Gibbs}, {Gillanders}, {Godambe}, {Grube}, {Guenette},
  {Gyuk}, {Hanna}, {Holder}, {Horan}, {Hui}, {Humensky}, {Imran}, {Kaaret},
  {Karlsson}, {Kertzman}, {Kieda}, {Konopelko}, {Krawczynski}, {Krennrich},
  {Lang}, {Le Bohec}, {Maier}, {McCann}, {McCutcheon}, {Millis}, {Moriarty},
  {Mukherjee}, {Ong}, {Otte}, {Pandel}, {Perkins}, {Pohl}, {Quinn}, {Ragan},
  {Reynolds}, {Roache}, {Rose}, {Schroedter}, {Sembroski}, {Smith}, {Steele},
  {Swordy}, {Theiling}, {Toner}, {Varlotta}, {Vassiliev}, {Vincent}, {Wagner},
  {Wakely}, {Ward}, {Weekes}, {Weinstein}, {Weisgarber}, {Williams}, {Wissel},
  {Wood}, {Zitzer}, {Kataoka}, {Cavazzuti}, {Cheung}, {Lott}, {Thompson}, \&
  {Tosti}}]{2009ApJ...706L.275A}
{Acciari}, V.~A., {Aliu}, E., {Arlen}, T., {et~al.} 2009, \apjl, 706, L275

\bibitem[{{Ahnen} {et~al.}(2016){Ahnen}, {Ansoldi}, {Antonelli}, {Antoranz},
  {Babic}, {Banerjee}, {Bangale}, {Barres de Almeida}, {Barrio}, {Becerra
  Gonz{\'a}lez}, {Bednarek}, {Bernardini}, {Biasuzzi}, {Biland}, {Blanch},
  {Bonnefoy}, {Bonnoli}, {Borracci}, {Bretz}, {Buson}, {Carmona}, {Carosi},
  {Chatterjee}, {Clavero}, {Colin}, {Colombo}, {Contreras}, {Cortina},
  {Covino}, {Da Vela}, {Dazzi}, {De Angelis}, {De Lotto}, {de O{\~n}a
  Wilhelmi}, {Delgado Mendez}, {Di Pierro}, {Dom{\'\i}nguez}, {Dominis
  Prester}, {Dorner}, {Doro}, {Einecke}, {Eisenacher Glawion}, {Elsaesser},
  {Fern{\'a}ndez-Barral}, {Fidalgo}, {Fonseca}, {Font}, {Frantzen}, {Fruck},
  {Galindo}, {Garc{\'\i}a L{\'o}pez}, {Garczarczyk}, {Garrido Terrats}, {Gaug},
  {Giammaria}, {Godinovi{\'c}}, {Gonz{\'a}lez Mu{\~n}oz}, {Gora}, {Guberman},
  {Hadasch}, {Hahn}, {Hanabata}, {Hayashida}, {Herrera}, {Hose}, {Hrupec},
  {Hughes}, {Idec}, {Kodani}, {Konno}, {Kubo}, {Kushida}, {La Barbera},
  {Lelas}, {Lindfors}, {Lombardi}, {Longo}, {L{\'o}pez}, {L{\'o}pez-Coto},
  {Lorenz}, {Majumdar}, {Makariev}, {Mallot}, {Maneva}, {Manganaro},
  {Mannheim}, {Maraschi}, {Marcote}, {Mariotti}, {Mart{\'\i}nez}, {Mazin},
  {Menzel}, {Miranda}, {Mirzoyan}, {Moralejo}, {Moretti}, {Nakajima},
  {Neustroev}, {Niedzwiecki}, {Nievas Rosillo}, {Nilsson}, {Nishijima}, {Noda},
  {Orito}, {Overkemping}, {Paiano}, {Palacio}, {Palatiello}, {Paneque},
  {Paoletti}, {Paredes}, {Paredes-Fortuny}, {Pedaletti}, {Persic}, {Poutanen},
  {Prada Moroni}, {Prandini}, {Puljak}, {Rhode}, {Rib{\'o}}, {Rico}, {Rodriguez
  Garcia}, {Saito}, {Satalecka}, {Schultz}, {Schweizer}, {Sillanp{\"a}{\"a}},
  {Sitarek}, {Snidaric}, {Sobczynska}, {Stamerra}, {Steinbring}, {Strzys},
  {Takalo}, {Takami}, {Tavecchio}, {Temnikov}, {Terzi{\'c}}, {Tescaro},
  {Teshima}, {Thaele}, {Torres}, {Toyama}, {Treves}, {Vazquez Acosta},
  {Verguilov}, {Vovk}, {Ward}, {Will}, {Wu}, {Zanin}, {Pfrommer}, {Pinzke}, \&
  {Zandanel}}]{2016A&A...589A..33A}
{Ahnen}, M.~L., {Ansoldi}, S., {Antonelli}, L.~A., {et~al.} 2016, \aap, 589,
  A33

\bibitem[{{Aleksi{\'c}} {et~al.}(2012){Aleksi{\'c}}, {Alvarez}, {Antonelli},
  {Antoranz}, {Asensio}, {Backes}, {Barres de Almeida}, {Barrio}, {Bastieri},
  {Becerra Gonz{\'a}lez}, {Bednarek}, {Berdyugin}, {Berger}, {Bernardini},
  {Biland}, {Blanch}, {Bock}, {Boller}, {Bonnoli}, {Borla Tridon}, {Braun},
  {Bretz}, {Ca{\~n}ellas}, {Carmona}, {Carosi}, {Colin}, {Colombo},
  {Contreras}, {Cortina}, {Cossio}, {Covino}, {Dazzi}, {de Angelis}, {de
  Caneva}, {de Cea Del Pozo}, {de Lotto}, {Delgado Mendez}, {Diago Ortega},
  {Doert}, {Dom{\'\i}nguez}, {Dominis Prester}, {Dorner}, {Doro}, {Eisenacher},
  {Elsaesser}, {Ferenc}, {Fonseca}, {Font}, {Fruck}, {Garc{\'\i}a L{\'o}pez},
  {Garczarczyk}, {Garrido}, {Giavitto}, {Godinovi{\'c}}, {Gozzini}, {Hadasch},
  {H{\"a}fner}, {Herrero}, {Hildebrand}, {H{\"o}hne-M{\"o}nch}, {Hose},
  {Hrupec}, {Jogler}, {Kellermann}, {Klepser}, {Kr{\"a}henb{\"u}hl}, {Krause},
  {Kushida}, {La Barbera}, {Lelas}, {Leonardo}, {Lewandowska}, {Lindfors},
  {Lombardi}, {L{\'o}pez}, {L{\'o}pez}, {L{\'o}pez-Oramas}, {Lorenz},
  {Makariev}, {Maneva}, {Mankuzhiyil}, {Mannheim}, {Maraschi}, {Mariotti},
  {Mart{\'\i}nez}, {Mazin}, {Meucci}, {Miranda}, {Mirzoyan}, {Mold{\'o}n},
  {Moralejo}, {Munar-Adrover}, {Niedzwiecki}, {Nieto}, {Nilsson}, {Nowak},
  {Orito}, {Paiano}, {Paneque}, {Paoletti}, {Pardo}, {Paredes}, {Partini},
  {Perez-Torres}, {Persic}, {Peruzzo}, {Pilia}, {Pochon}, {Prada}, {Prada
  Moroni}, {Prandini}, {Puerto Gimenez}, {Puljak}, {Reichardt}, {Reinthal},
  {Rhode}, {Rib{\'o}}, {Rico}, {R{\"u}gamer}, {Saggion}, {Saito}, {Saito},
  {Salvati}, {Satalecka}, {Scalzotto}, {Scapin}, {Schultz}, {Schweizer},
  {Shayduk}, {Shore}, {Sillanp{\"a}{\"a}}, {Sitarek}, {Snidaric}, {Sobczynska},
  {Spanier}, {Spiro}, {Stamatescu}, {Stamerra}, {Steinke}, {Storz}, {Strah},
  {Sun}, {Suri{\'c}}, {Takalo}, {Takami}, {Tavecchio}, {Temnikov},
  {Terzi{\'c}}, {Tescaro}, {Teshima}, {Tibolla}, {Torres}, {Treves},
  {Uellenbeck}, {Vankov}, {Vogler}, {Wagner}, {Weitzel}, {Zabalza}, {Zandanel},
  {Zanin}, {MAGIC Collaboration}, {Pfrommer}, \&
  {Pinzke}}]{2012A&A...541A..99A}
{Aleksi{\'c}}, J., {Alvarez}, E.~A., {Antonelli}, L.~A., {et~al.} 2012, \aap,
  541, A99

\bibitem[{{Aleksi{\'c}} {et~al.}(2010){Aleksi{\'c}}, {Antonelli}, {Antoranz},
  {Backes}, {Baixeras}, {Balestra}, {Barrio}, {Bastieri}, {Becerra
  Gonz{\'a}lez}, {Bednarek}, {Berdyugin}, {Berger}, {Bernardini}, {Biland},
  {Bock}, {Bonnoli}, {Bordas}, {Borla Tridon}, {Bosch-Ramon}, {Bose}, {Braun},
  {Bretz}, {Britzger}, {Camara}, {Carmona}, {Carosi}, {Colin}, {Commichau},
  {Contreras}, {Cortina}, {Costado}, {Covino}, {Dazzi}, {De Angelis}, {De Cea
  del Pozo}, {De los Reyes}, {De Lotto}, {De Maria}, {De Sabata}, {Delgado
  Mendez}, {Doert}, {Dom{\'\i}nguez}, {Dominis Prester}, {Dorner}, {Doro},
  {Elsaesser}, {Errando}, {Ferenc}, {Fonseca}, {Font}, {Galante}, {Garc{\'\i}a
  L{\'o}pez}, {Garczarczyk}, {Gaug}, {Godinovic}, {Hadasch}, {Herrero},
  {Hildebrand}, {H{\"o}hne-M{\"o}nch}, {Hose}, {Hrupec}, {Hsu}, {Jogler},
  {Klepser}, {Kr{\"a}henb{\"u}hl}, {Kranich}, {La Barbera}, {Laille},
  {Leonardo}, {Lindfors}, {Lombardi}, {Longo}, {L{\'o}pez}, {Lorenz},
  {Majumdar}, {Maneva}, {Mankuzhiyil}, {Mannheim}, {Maraschi}, {Mariotti},
  {Mart{\'\i}nez}, {Mazin}, {Meucci}, {Miranda}, {Mirzoyan}, {Miyamoto},
  {Mold{\'o}n}, {Moles}, {Moralejo}, {Nieto}, {Nilsson}, {Ninkovic}, {Orito},
  {Oya}, {Paiano}, {Paoletti}, {Paredes}, {Partini}, {Pasanen}, {Pascoli},
  {Pauss}, {Pegna}, {Perez-Torres}, {Persic}, {Peruzzo}, {Prada}, {Prandini},
  {Puchades}, {Puljak}, {Reichardt}, {Rhode}, {Rib{\'o}}, {Rico}, {Rissi},
  {R{\"u}gamer}, {Saggion}, {Saito}, {Salvati}, {S{\'a}nchez-Conde},
  {Satalecka}, {Scalzotto}, {Scapin}, {Schultz}, {Schweizer}, {Shayduk},
  {Shore}, {Sierpowska-Bartosik}, {Sillanp{\"a}{\"a}}, {Sitarek}, {Sobczynska},
  {Spanier}, {Spiro}, {Stamerra}, {Steinke}, {Struebig}, {Suric}, {Takalo},
  {Tavecchio}, {Temnikov}, {Terzic}, {Tescaro}, {Teshima}, {Torres}, {Vankov},
  {Wagner}, {Zabalza}, {Zandanel}, {Zanin}, {Zapatero}, {Pfrommer}, {Pinzke},
  {En{\ss}lin}, {Inoue}, {Ghisellini}, \& {MAGIC
  Collaboration}}]{2010ApJ...710..634A}
{Aleksi{\'c}}, J., {Antonelli}, L.~A., {Antoranz}, P., {et~al.} 2010, \apj,
  710, 634

\bibitem[{{Astropy Collaboration} {et~al.}(2022){Astropy Collaboration},
  {Price-Whelan}, {Lim}, {Earl}, {Starkman}, {Bradley}, {Shupe}, {Patil},
  {Corrales}, {Brasseur}, {N{\"o}the}, {Donath}, {Tollerud}, {Morris},
  {Ginsburg}, {Vaher}, {Weaver}, {Tocknell}, {Jamieson}, {van Kerkwijk},
  {Robitaille}, {Merry}, {Bachetti}, {G{\"u}nther}, {Aldcroft},
  {Alvarado-Montes}, {Archibald}, {B{\'o}di}, {Bapat}, {Barentsen},
  {Baz{\'a}n}, {Biswas}, {Boquien}, {Burke}, {Cara}, {Cara}, {Conroy},
  {Conseil}, {Craig}, {Cross}, {Cruz}, {D'Eugenio}, {Dencheva}, {Devillepoix},
  {Dietrich}, {Eigenbrot}, {Erben}, {Ferreira}, {Foreman-Mackey}, {Fox},
  {Freij}, {Garg}, {Geda}, {Glattly}, {Gondhalekar}, {Gordon}, {Grant},
  {Greenfield}, {Groener}, {Guest}, {Gurovich}, {Handberg}, {Hart},
  {Hatfield-Dodds}, {Homeier}, {Hosseinzadeh}, {Jenness}, {Jones}, {Joseph},
  {Kalmbach}, {Karamehmetoglu}, {Ka{\l}uszy{\'n}ski}, {Kelley}, {Kern},
  {Kerzendorf}, {Koch}, {Kulumani}, {Lee}, {Ly}, {Ma}, {MacBride}, {Maljaars},
  {Muna}, {Murphy}, {Norman}, {O'Steen}, {Oman}, {Pacifici}, {Pascual},
  {Pascual-Granado}, {Patil}, {Perren}, {Pickering}, {Rastogi}, {Roulston},
  {Ryan}, {Rykoff}, {Sabater}, {Sakurikar}, {Salgado}, {Sanghi}, {Saunders},
  {Savchenko}, {Schwardt}, {Seifert-Eckert}, {Shih}, {Jain}, {Shukla}, {Sick},
  {Simpson}, {Singanamalla}, {Singer}, {Singhal}, {Sinha}, {Sip{\H{o}}cz},
  {Spitler}, {Stansby}, {Streicher}, {{\v{S}}umak}, {Swinbank}, {Taranu},
  {Tewary}, {Tremblay}, {de Val-Borro}, {Van Kooten}, {Vasovi{\'c}}, {Verma},
  {de Miranda Cardoso}, {Williams}, {Wilson}, {Winkel}, {Wood-Vasey}, {Xue},
  {Yoachim}, {Zhang}, {Zonca}, \& {Astropy Project
  Contributors}}]{2022ApJ...935..167A}
{Astropy Collaboration}, {Price-Whelan}, A.~M., {Lim}, P.~L., {et~al.} 2022,
  \apj, 935, 167

\bibitem[{{Astropy Collaboration} {et~al.}(2018){Astropy Collaboration},
  {Price-Whelan}, {Sip{\H{o}}cz}, {G{\"u}nther}, {Lim}, {Crawford}, {Conseil},
  {Shupe}, {Craig}, {Dencheva}, {Ginsburg}, {VanderPlas}, {Bradley},
  {P{\'e}rez-Su{\'a}rez}, {de Val-Borro}, {Aldcroft}, {Cruz}, {Robitaille},
  {Tollerud}, {Ardelean}, {Babej}, {Bach}, {Bachetti}, {Bakanov}, {Bamford},
  {Barentsen}, {Barmby}, {Baumbach}, {Berry}, {Biscani}, {Boquien}, {Bostroem},
  {Bouma}, {Brammer}, {Bray}, {Breytenbach}, {Buddelmeijer}, {Burke},
  {Calderone}, {Cano Rodr{\'\i}guez}, {Cara}, {Cardoso}, {Cheedella}, {Copin},
  {Corrales}, {Crichton}, {D'Avella}, {Deil}, {Depagne}, {Dietrich}, {Donath},
  {Droettboom}, {Earl}, {Erben}, {Fabbro}, {Ferreira}, {Finethy}, {Fox},
  {Garrison}, {Gibbons}, {Goldstein}, {Gommers}, {Greco}, {Greenfield},
  {Groener}, {Grollier}, {Hagen}, {Hirst}, {Homeier}, {Horton}, {Hosseinzadeh},
  {Hu}, {Hunkeler}, {Ivezi{\'c}}, {Jain}, {Jenness}, {Kanarek}, {Kendrew},
  {Kern}, {Kerzendorf}, {Khvalko}, {King}, {Kirkby}, {Kulkarni}, {Kumar},
  {Lee}, {Lenz}, {Littlefair}, {Ma}, {Macleod}, {Mastropietro}, {McCully},
  {Montagnac}, {Morris}, {Mueller}, {Mumford}, {Muna}, {Murphy}, {Nelson},
  {Nguyen}, {Ninan}, {N{\"o}the}, {Ogaz}, {Oh}, {Parejko}, {Parley}, {Pascual},
  {Patil}, {Patil}, {Plunkett}, {Prochaska}, {Rastogi}, {Reddy Janga},
  {Sabater}, {Sakurikar}, {Seifert}, {Sherbert}, {Sherwood-Taylor}, {Shih},
  {Sick}, {Silbiger}, {Singanamalla}, {Singer}, {Sladen}, {Sooley},
  {Sornarajah}, {Streicher}, {Teuben}, {Thomas}, {Tremblay}, {Turner},
  {Terr{\'o}n}, {van Kerkwijk}, {de la Vega}, {Watkins}, {Weaver}, {Whitmore},
  {Woillez}, {Zabalza}, \& {Astropy Contributors}}]{2018AJ....156..123A}
{Astropy Collaboration}, {Price-Whelan}, A.~M., {Sip{\H{o}}cz}, B.~M., {et~al.}
  2018, \aj, 156, 123

\bibitem[{{Astropy Collaboration} {et~al.}(2013){Astropy Collaboration},
  {Robitaille}, {Tollerud}, {Greenfield}, {Droettboom}, {Bray}, {Aldcroft},
  {Davis}, {Ginsburg}, {Price-Whelan}, {Kerzendorf}, {Conley}, {Crighton},
  {Barbary}, {Muna}, {Ferguson}, {Grollier}, {Parikh}, {Nair}, {Unther},
  {Deil}, {Woillez}, {Conseil}, {Kramer}, {Turner}, {Singer}, {Fox}, {Weaver},
  {Zabalza}, {Edwards}, {Azalee Bostroem}, {Burke}, {Casey}, {Crawford},
  {Dencheva}, {Ely}, {Jenness}, {Labrie}, {Lim}, {Pierfederici}, {Pontzen},
  {Ptak}, {Refsdal}, {Servillat}, \& {Streicher}}]{2013A&A...558A..33A}
{Astropy Collaboration}, {Robitaille}, T.~P., {Tollerud}, E.~J., {et~al.} 2013,
  \aap, 558, A33

\bibitem[{{Biava} {et~al.}(2024){Biava}, {Bonafede}, {Gastaldello}, {Botteon},
  {Brienza}, {Shimwell}, {Brunetti}, {Bruno}, {Rajpurohit}, {Riseley}, {van
  Weeren}, {Rossetti}, {Cassano}, {De Gasperin}, {Drabent}, {Rottgering},
  {Edge}, \& {Tasse}}]{2024A&A...686A..82B}
{Biava}, N., {Bonafede}, A., {Gastaldello}, F., {et~al.} 2024, \aap, 686, A82

\bibitem[{{Biava} {et~al.}(2021){Biava}, {de Gasperin}, {Bonafede}, {Edler},
  {Giacintucci}, {Mazzotta}, {Brunetti}, {Botteon}, {Br{\"u}ggen}, {Cassano},
  {Drabent}, {Edge}, {En{\ss}lin}, {Gastaldello}, {Riseley}, {Rossetti},
  {Rottgering}, {Shimwell}, {Tasse}, \& {van Weeren}}]{2021MNRAS.508.3995B}
{Biava}, N., {de Gasperin}, F., {Bonafede}, A., {et~al.} 2021, \mnras, 508,
  3995

\bibitem[{{B{\^\i}rzan} {et~al.}(2020){B{\^\i}rzan}, {Rafferty}, {Br{\"u}ggen},
  {Botteon}, {Brunetti}, {Cuciti}, {Edge}, {Morganti}, {R{\"o}ttgering}, \&
  {Shimwell}}]{2020MNRAS.496.2613B}
{B{\^\i}rzan}, L., {Rafferty}, D.~A., {Br{\"u}ggen}, M., {et~al.} 2020, \mnras,
  496, 2613

\bibitem[{{Blundell} {et~al.}(2002){Blundell}, {Kassim}, \&
  {Perley}}]{2002IAUS..199..189B}
{Blundell}, K.~M., {Kassim}, N.~E., \& {Perley}, R.~A. 2002, in IAU Symposium,
  Vol. 199, The Universe at Low Radio Frequencies, ed. A.~{Pramesh Rao},
  G.~{Swarup}, \& {Gopal-Krishna}, 189

\bibitem[{{Boehringer} {et~al.}(1993){Boehringer}, {Voges}, {Fabian}, {Edge},
  \& {Neumann}}]{1993MNRAS.264L..25B}
{Boehringer}, H., {Voges}, W., {Fabian}, A.~C., {Edge}, A.~C., \& {Neumann},
  D.~M. 1993, \mnras, 264, L25

\bibitem[{{Bonafede} {et~al.}(2014){Bonafede}, {Intema}, {Bruggen}, {Russell},
  {Ogrean}, {Basu}, {Sommer}, {van Weeren}, {Cassano}, {Fabian}, \&
  {Rottgering}}]{2014MNRAS.444L..44B}
{Bonafede}, A., {Intema}, H.~T., {Bruggen}, M., {et~al.} 2014, \mnras, 444, L44

\bibitem[{{Boxelaar} {et~al.}(2025){Boxelaar}, {De Gasperin}, {Hardcastle},
  {Croston}, {Morabito}, {van Weeren}, \& {Edler}}]{2025A&A...704A..65B}
{Boxelaar}, J.~M., {De Gasperin}, F., {Hardcastle}, M.~J., {et~al.} 2025, \aap,
  704, A65

\bibitem[{{Branduardi-Raymont} {et~al.}(1981){Branduardi-Raymont}, {Fabricant},
  {Feigelson}, {Gorenstein}, {Grindlay}, {Soltan}, \&
  {Zamorani}}]{1981ApJ...248...55B}
{Branduardi-Raymont}, G., {Fabricant}, D., {Feigelson}, E., {et~al.} 1981,
  \apj, 248, 55

\bibitem[{{Braude}(1966)}]{1966Natur.210Q..80B}
{Braude}, S.~Y. 1966, \nat, 210, 80

\bibitem[{{Brentjens}(2011)}]{2011A&A...526A...9B}
{Brentjens}, M.~A. 2011, \aap, 526, A9

\bibitem[{{Br{\"u}ggen}(2003)}]{2003ApJ...592..839B}
{Br{\"u}ggen}, M. 2003, \apj, 592, 839

\bibitem[{{Br{\"u}ggen} \& {Kaiser}(2002)}]{2002Natur.418..301B}
{Br{\"u}ggen}, M. \& {Kaiser}, C.~R. 2002, \nat, 418, 301

\bibitem[{{Bruno} {et~al.}(2023){Bruno}, {Botteon}, {Shimwell}, {Cuciti}, {de
  Gasperin}, {Brunetti}, {Dallacasa}, {Gastaldello}, {Rossetti}, {van Weeren},
  {Venturi}, {Russo}, {Taffoni}, {Cassano}, {Biava}, {Lusetti}, {Bonafede},
  {Ghizzardi}, \& {De Grandi}}]{2023A&A...678A.133B}
{Bruno}, L., {Botteon}, A., {Shimwell}, T., {et~al.} 2023, \aap, 678, A133

\bibitem[{{Buote}(2001)}]{2001ApJ...553L..15B}
{Buote}, D.~A. 2001, \apjl, 553, L15

\bibitem[{{Burns} {et~al.}(1992){Burns}, {Sulkanen}, {Gisler}, \&
  {Perley}}]{1992ApJ...388L..49B}
{Burns}, J.~O., {Sulkanen}, M.~E., {Gisler}, G.~R., \& {Perley}, R.~A. 1992,
  \apjl, 388, L49

\bibitem[{{Carilli} {et~al.}(1994){Carilli}, {Perley}, \&
  {Harris}}]{1994MNRAS.270..173C}
{Carilli}, C.~L., {Perley}, R.~A., \& {Harris}, D.~E. 1994, \mnras, 270, 173

\bibitem[{{Cassano} {et~al.}(2010){Cassano}, {Ettori}, {Giacintucci},
  {Brunetti}, {Markevitch}, {Venturi}, \& {Gitti}}]{2010ApJ...721L..82C}
{Cassano}, R., {Ettori}, S., {Giacintucci}, S., {et~al.} 2010, \apjl, 721, L82

\bibitem[{{Cassano} {et~al.}(2008){Cassano}, {Gitti}, \&
  {Brunetti}}]{2008A&A...486L..31C}
{Cassano}, R., {Gitti}, M., \& {Brunetti}, G. 2008, \aap, 486, L31

\bibitem[{{Churazov} {et~al.}(2001){Churazov}, {Br{\"u}ggen}, {Kaiser},
  {B{\"o}hringer}, \& {Forman}}]{2001ApJ...554..261C}
{Churazov}, E., {Br{\"u}ggen}, M., {Kaiser}, C.~R., {B{\"o}hringer}, H., \&
  {Forman}, W. 2001, \apj, 554, 261

\bibitem[{{Churazov} {et~al.}(2000){Churazov}, {Forman}, {Jones}, \&
  {B{\"o}hringer}}]{2000A&A...356..788C}
{Churazov}, E., {Forman}, W., {Jones}, C., \& {B{\"o}hringer}, H. 2000, \aap,
  356, 788

\bibitem[{{Clarke} {et~al.}(2005){Clarke}, {Sarazin}, {Blanton}, {Neumann}, \&
  {Kassim}}]{2005ApJ...625..748C}
{Clarke}, T.~E., {Sarazin}, C.~L., {Blanton}, E.~L., {Neumann}, D.~M., \&
  {Kassim}, N.~E. 2005, \apj, 625, 748

\bibitem[{{Condon} {et~al.}(1998){Condon}, {Cotton}, {Greisen}, {Yin},
  {Perley}, {Taylor}, \& {Broderick}}]{1998AJ....115.1693C}
{Condon}, J.~J., {Cotton}, W.~D., {Greisen}, E.~W., {et~al.} 1998, \aj, 115,
  1693

\bibitem[{{de Gasperin} {et~al.}(2019){de Gasperin}, {Dijkema}, {Drabent},
  {Mevius}, {Rafferty}, {van Weeren}, {Br{\"u}ggen}, {Callingham}, {Emig},
  {Heald}, {Intema}, {Morabito}, {Offringa}, {Oonk}, {Orr{\`u}},
  {R{\"o}ttgering}, {Sabater}, {Shimwell}, {Shulevski}, \&
  {Williams}}]{2019A&A...622A...5D}
{de Gasperin}, F., {Dijkema}, T.~J., {Drabent}, A., {et~al.} 2019, \aap, 622,
  A5

\bibitem[{{de Gasperin} {et~al.}(2023){de Gasperin}, {Edler}, {Williams},
  {Callingham}, {Asabere}, {Br{\"u}ggen}, {Brunetti}, {Dijkema}, {Hardcastle},
  {Iacobelli}, {Offringa}, {Norden}, {R{\"o}ttgering}, {Shimwell}, {van
  Weeren}, {Tasse}, {Bomans}, {Bonafede}, {Botteon}, {Cassano}, {Chy{\.z}y},
  {Cuciti}, {Emig}, {Kadler}, {Miley}, {Mingo}, {Oei}, {Prandoni}, {Schwarz},
  \& {Zarka}}]{2023A&A...673A.165D}
{de Gasperin}, F., {Edler}, H.~W., {Williams}, W.~L., {et~al.} 2023, \aap, 673,
  A165

\bibitem[{{de Gasperin} {et~al.}(2021){de Gasperin}, {Williams}, {Best},
  {Br{\"u}ggen}, {Brunetti}, {Cuciti}, {Dijkema}, {Hardcastle}, {Norden},
  {Offringa}, {Shimwell}, {van Weeren}, {Bomans}, {Bonafede}, {Botteon},
  {Callingham}, {Cassano}, {Chy{\.z}y}, {Emig}, {Edler}, {Haverkorn}, {Heald},
  {Heesen}, {Iacobelli}, {Intema}, {Kadler}, {Ma{\l}ek}, {Mevius}, {Miley},
  {Mingo}, {Morabito}, {Sabater}, {Morganti}, {Orr{\'u}}, {Pizzo}, {Prandoni},
  {Shulevski}, {Tasse}, {Vaccari}, {Zarka}, \&
  {R{\"o}ttgering}}]{2021A&A...648A.104D}
{de Gasperin}, F., {Williams}, W.~L., {Best}, P., {et~al.} 2021, \aap, 648,
  A104

\bibitem[{{Douglass} {et~al.}(2018){Douglass}, {Blanton}, {Randall}, {Clarke},
  {Edwards}, {Sabry}, \& {ZuHone}}]{2018ApJ...868..121D}
{Douglass}, E.~M., {Blanton}, E.~L., {Randall}, S.~W., {et~al.} 2018, \apj,
  868, 121

\bibitem[{{Fabian}(2012)}]{2012ARA&A..50..455F}
{Fabian}, A.~C. 2012, \araa, 50, 455

\bibitem[{{Fabian} {et~al.}(2002){Fabian}, {Celotti}, {Blundell}, {Kassim}, \&
  {Perley}}]{2002MNRAS.331..369F}
{Fabian}, A.~C., {Celotti}, A., {Blundell}, K.~M., {Kassim}, N.~E., \&
  {Perley}, R.~A. 2002, \mnras, 331, 369

\bibitem[{{Fabian} {et~al.}(1981){Fabian}, {Hu}, {Cowie}, \&
  {Grindlay}}]{1981ApJ...248...47F}
{Fabian}, A.~C., {Hu}, E.~M., {Cowie}, L.~L., \& {Grindlay}, J. 1981, \apj,
  248, 47

\bibitem[{{Fabian} {et~al.}(2011){Fabian}, {Sanders}, {Allen}, {Canning},
  {Churazov}, {Crawford}, {Forman}, {Gabany}, {Hlavacek-Larrondo}, {Johnstone},
  {Russell}, {Reynolds}, {Salom{\'e}}, {Taylor}, \&
  {Young}}]{2011MNRAS.418.2154F}
{Fabian}, A.~C., {Sanders}, J.~S., {Allen}, S.~W., {et~al.} 2011, \mnras, 418,
  2154

\bibitem[{{Fabian} {et~al.}(2003){Fabian}, {Sanders}, {Allen}, {Crawford},
  {Iwasawa}, {Johnstone}, {Schmidt}, \& {Taylor}}]{2003MNRAS.344L..43F}
{Fabian}, A.~C., {Sanders}, J.~S., {Allen}, S.~W., {et~al.} 2003, \mnras, 344,
  L43

\bibitem[{{Fabian} {et~al.}(2000){Fabian}, {Sanders}, {Ettori}, {Taylor},
  {Allen}, {Crawford}, {Iwasawa}, {Johnstone}, \& {Ogle}}]{2000MNRAS.318L..65F}
{Fabian}, A.~C., {Sanders}, J.~S., {Ettori}, S., {et~al.} 2000, \mnras, 318,
  L65

\bibitem[{{Fabian} {et~al.}(2006){Fabian}, {Sanders}, {Taylor}, {Allen},
  {Crawford}, {Johnstone}, \& {Iwasawa}}]{2006MNRAS.366..417F}
{Fabian}, A.~C., {Sanders}, J.~S., {Taylor}, G.~B., {et~al.} 2006, \mnras, 366,
  417

\bibitem[{{Feretti} {et~al.}(2012){Feretti}, {Giovannini}, {Govoni}, \&
  {Murgia}}]{2012A&ARv..20...54F}
{Feretti}, L., {Giovannini}, G., {Govoni}, F., \& {Murgia}, M. 2012, \aapr, 20,
  54

\bibitem[{{Fujita} {et~al.}(2007){Fujita}, {Kohri}, {Yamazaki}, \&
  {Kino}}]{2007ApJ...663L..61F}
{Fujita}, Y., {Kohri}, K., {Yamazaki}, R., \& {Kino}, M. 2007, \apjl, 663, L61

\bibitem[{{Gendron-Marsolais} {et~al.}(2017){Gendron-Marsolais},
  {Hlavacek-Larrondo}, {van Weeren}, {Clarke}, {Fabian}, {Intema}, {Taylor},
  {Blundell}, \& {Sanders}}]{2017MNRAS.469.3872G}
{Gendron-Marsolais}, M., {Hlavacek-Larrondo}, J., {van Weeren}, R.~J., {et~al.}
  2017, \mnras, 469, 3872

\bibitem[{{Gendron-Marsolais} {et~al.}(2020){Gendron-Marsolais},
  {Hlavacek-Larrondo}, {van Weeren}, {Rudnick}, {Clarke}, {Sebastian},
  {Mroczkowski}, {Fabian}, {Blundell}, {Sheldahl}, {Nyland}, {Sanders},
  {Peters}, \& {Intema}}]{2020MNRAS.499.5791G}
{Gendron-Marsolais}, M., {Hlavacek-Larrondo}, J., {van Weeren}, R.~J., {et~al.}
  2020, \mnras, 499, 5791

\bibitem[{{Gendron-Marsolais} {et~al.}(2021){Gendron-Marsolais}, {Hull},
  {Perley}, {Rudnick}, {Kraft}, {Hlavacek-Larrondo}, {Fabian}, {Roediger}, {van
  Weeren}, {Richard-Laferri{\`e}re}, {Golden-Marx}, {Arakawa}, \&
  {McBride}}]{2021ApJ...911...56G}
{Gendron-Marsolais}, M.~L., {Hull}, C.~L.~H., {Perley}, R., {et~al.} 2021,
  \apj, 911, 56

\bibitem[{{Giacintucci} {et~al.}(2017){Giacintucci}, {Markevitch}, {Cassano},
  {Venturi}, {Clarke}, \& {Brunetti}}]{2017ApJ...841...71G}
{Giacintucci}, S., {Markevitch}, M., {Cassano}, R., {et~al.} 2017, \apj, 841,
  71

\bibitem[{{Giacintucci} {et~al.}(2019){Giacintucci}, {Markevitch}, {Cassano},
  {Venturi}, {Clarke}, {Kale}, \& {Cuciti}}]{2019ApJ...880...70G}
{Giacintucci}, S., {Markevitch}, M., {Cassano}, R., {et~al.} 2019, \apj, 880,
  70

\bibitem[{{Giacintucci} {et~al.}(2014){Giacintucci}, {Markevitch}, {Venturi},
  {Clarke}, {Cassano}, \& {Mazzotta}}]{2014ApJ...781....9G}
{Giacintucci}, S., {Markevitch}, M., {Venturi}, T., {et~al.} 2014, \apj, 781, 9

\bibitem[{{Gitti} {et~al.}(2004){Gitti}, {Brunetti}, {Feretti}, \&
  {Setti}}]{2004A&A...417....1G}
{Gitti}, M., {Brunetti}, G., {Feretti}, L., \& {Setti}, G. 2004, \aap, 417, 1

\bibitem[{{Gitti} {et~al.}(2002){Gitti}, {Brunetti}, \&
  {Setti}}]{2002A&A...386..456G}
{Gitti}, M., {Brunetti}, G., \& {Setti}, G. 2002, \aap, 386, 456

\bibitem[{{Groeneveld} {et~al.}(2022){Groeneveld}, {van Weeren}, {Miley},
  {Morabito}, {de Gasperin}, {Callingham}, {Sweijen}, {Br{\"u}ggen}, {Botteon},
  {Offringa}, {Brunetti}, {Moldon}, {Bondi}, {Kappes}, \&
  {R{\"o}ttgering}}]{2022A&A...658A...9G}
{Groeneveld}, C., {van Weeren}, R.~J., {Miley}, G.~K., {et~al.} 2022, \aap,
  658, A9

\bibitem[{{Groeneveld} {et~al.}(2024){Groeneveld}, {van Weeren}, {Osinga},
  {Williams}, {Callingham}, {de Gasperin}, {Botteon}, {Shimwell}, {Sweijen},
  {de Jong}, {Jansen}, {Miley}, {Brunetti}, {Br{\"u}ggen}, \&
  {R{\"o}ttgering}}]{2024NatAs...8..786G}
{Groeneveld}, C., {van Weeren}, R.~J., {Osinga}, E., {et~al.} 2024, Nature
  Astronomy, 8, 786

\bibitem[{{Harris} \& {Mulholland}(2017)}]{2017ApJ...839..102H}
{Harris}, W.~E. \& {Mulholland}, C.~J. 2017, \apj, 839, 102

\bibitem[{{HyeongHan} {et~al.}(2025){HyeongHan}, {Jee}, {Lee}, {ZuHone},
  {Zhuravleva}, {Kang}, \& {Hwang}}]{2025NatAs...9..925H}
{HyeongHan}, K., {Jee}, M.~J., {Lee}, W., {et~al.} 2025, Nature Astronomy, 9,
  925

\bibitem[{{Intema} {et~al.}(2017){Intema}, {Jagannathan}, {Mooley}, \&
  {Frail}}]{2017A&A...598A..78I}
{Intema}, H.~T., {Jagannathan}, P., {Mooley}, K.~P., \& {Frail}, D.~A. 2017,
  \aap, 598, A78

\bibitem[{{Irfan} \& {Puglisi}(2025)}]{2025arXiv251208522I}
{Irfan}, M.~O. \& {Puglisi}, G. 2025, arXiv e-prints, arXiv:2512.08522

\bibitem[{{Kassim} {et~al.}(2007){Kassim}, {Lazio}, {Erickson}, {Perley},
  {Cotton}, {Greisen}, {Cohen}, {Hicks}, {Schmitt}, \&
  {Katz}}]{2007ApJS..172..686K}
{Kassim}, N.~E., {Lazio}, T. J.~W., {Erickson}, W.~C., {et~al.} 2007, \apjs,
  172, 686

\bibitem[{{Keshet}(2024)}]{2024MNRAS.527.1194K}
{Keshet}, U. 2024, \mnras, 527, 1194

\bibitem[{{Kluge} {et~al.}(2025){Kluge}, {Hatch}, {Montes}, {Golden-Marx},
  {Gonzalez}, {Cuillandre}, {Bolzonella}, {Lan{\c{c}}on}, {Laureijs},
  {Saifollahi}, {Schirmer}, {Stone}, {Boselli}, {Cantiello}, {Sorce},
  {Marleau}, {Duc}, {Sola}, {Urbano}, {Ahad}, {Bah{\'e}}, {Bamford},
  {Bellhouse}, {Buitrago}, {Dimauro}, {Durret}, {Ellien}, {Jimenez-Teja},
  {Slezak}, {Aghanim}, {Altieri}, {Andreon}, {Auricchio}, {Baldi}, {Balestra},
  {Bardelli}, {Bender}, {Bonino}, {Branchini}, {Brescia}, {Brinchmann},
  {Camera}, {Candini}, {Capobianco}, {Carbone}, {Carretero}, {Casas},
  {Castellano}, {Cavuoti}, {Cimatti}, {Congedo}, {Conselice}, {Conversi},
  {Copin}, {Courbin}, {Courtois}, {Cropper}, {Da Silva}, {Degaudenzi}, {Dinis},
  {Duncan}, {Dupac}, {Dusini}, {Farina}, {Farrens}, {Ferriol}, {Fosalba},
  {Frailis}, {Franceschi}, {Fumana}, {Galeotta}, {Garilli}, {Gillard},
  {Gillis}, {Giocoli}, {G{\'o}mez-Alvarez}, {Granett}, {Grazian}, {Grupp},
  {Guzzo}, {Haugan}, {Hoar}, {Hoekstra}, {Holmes}, {Hook}, {Hormuth},
  {Hornstrup}, {Hudelot}, {Jahnke}, {Keih{\"a}nen}, {Kermiche}, {Kiessling},
  {Kitching}, {Kohley}, {Kubik}, {K{\"u}mmel}, {Kunz}, {Kurki-Suonio}, {Lahav},
  {Ligori}, {Lilje}, {Lindholm}, {Lloro}, {Maiorano}, {Mansutti}, {Marggraf},
  {Markovic}, {Martinet}, {Marulli}, {Massey}, {Maurogordato}, {McCracken},
  {Medinaceli}, {Mei}, {Melchior}, {Mellier}, {Meneghetti}, {Merlin}, {Meylan},
  {Moresco}, {Moscardini}, {Munari}, {Nichol}, {Niemi}, {Nightingale},
  {Padilla}, {Paltani}, {Pasian}, {Pedersen}, {Percival}, {Pettorino}, {Pires},
  {Polenta}, {Poncet}, {Popa}, {Pozzetti}, {Racca}, {Raison}, {Rebolo},
  {Renzi}, {Rhodes}, {Riccio}, {Rix}, {Romelli}, {Roncarelli}, {Rossetti},
  {Saglia}, {Sapone}, {Sartoris}, {Sauvage}, {Scaramella}, {Schneider},
  {Schrabback}, {Secroun}, {Seidel}, {Seiffert}, {Serrano}, {Sirignano},
  {Sirri}, {Skottfelt}, {Stanco}, {Tallada-Cresp{\'\i}}, {Taylor}, {Teplitz},
  {Tereno}, {Toledo-Moreo}, {Torradeflot}, {Tutusaus}, {Valentijn},
  {Valenziano}, {Vassallo}, {Verdoes Kleijn}, {Veropalumbo}, {Wang}, {Weller},
  {Williams}, {Zamorani}, {Zucca}, {Biviano}, {Burigana}, {De Lucia}, {George},
  {Scottez}, {Simon}, {Mora}, {Mart{\'\i}n-Fleitas}, {Ruppin}, \&
  {Scott}}]{2025A&A...697A..13K}
{Kluge}, M., {Hatch}, N.~A., {Montes}, M., {et~al.} 2025, \aap, 697, A13

\bibitem[{{Knowles} {et~al.}(2022){Knowles}, {Cotton}, {Rudnick}, {Camilo},
  {Goedhart}, {Deane}, {Ramatsoku}, {Bietenholz}, {Br{\"u}ggen}, {Button},
  {Chen}, {Chibueze}, {Clarke}, {de Gasperin}, {Ianjamasimanana}, {J{\'o}zsa},
  {Hilton}, {Kesebonye}, {Kolokythas}, {Kraan-Korteweg}, {Lawrie}, {Lochner},
  {Loubser}, {Marchegiani}, {Mhlahlo}, {Moodley}, {Murphy}, {Namumba},
  {Oozeer}, {Parekh}, {Pillay}, {Passmoor}, {Ramaila}, {Ranchod},
  {Retana-Montenegro}, {Sebokolodi}, {Sikhosana}, {Smirnov}, {Thorat},
  {Venturi}, {Abbott}, {Adam}, {Adams}, {Aldera}, {Bauermeister}, {Bennett},
  {Bode}, {Botha}, {Botha}, {Brederode}, {Buchner}, {Burger}, {Cheetham}, {de
  Villiers}, {Dikgale-Mahlakoana}, {du Toit}, {Esterhuyse}, {Fadana},
  {Fanaroff}, {Fataar}, {Foley}, {Fourie}, {Frank}, {Gamatham}, {Gatsi},
  {Geyer}, {Gouws}, {Gumede}, {Heywood}, {Hlakola}, {Hokwana}, {Hoosen},
  {Horn}, {Horrell}, {Hugo}, {Isaacson}, {Jonas}, {Jordaan}, {Joubert},
  {Julie}, {Kapp}, {Kasper}, {Kenyon}, {Kotz{\'e}}, {Kotze}, {Kriek}, {Kriel},
  {Krishnan}, {Kusel}, {Legodi}, {Lehmensiek}, {Liebenberg}, {Lord}, {Lunsky},
  {Madisa}, {Magnus}, {Main}, {Makhaba}, {Makhathini}, {Malan}, {Manley},
  {Marais}, {Maree}, {Martens}, {Mauch}, {McAlpine}, {Merry}, {Millenaar},
  {Mokone}, {Monama}, {Mphego}, {New}, {Ngcebetsha}, {Ngoasheng}, {Ockards},
  {Otto}, {Patel}, {Peens-Hough}, {Perkins}, {Ramanujam}, {Ramudzuli},
  {Ratcliffe}, {Renil}, {Robyntjies}, {Rust}, {Salie}, {Sambu}, {Schollar},
  {Schwardt}, {Schwartz}, {Serylak}, {Siebrits}, {Sirothia}, {Slabber},
  {Sofeya}, {Taljaard}, {Tasse}, {Tiplady}, {Toruvanda}, {Twum}, {van Balla},
  {van der Byl}, {van der Merwe}, {van Dyk}, {Van Tonder}, {Van Wyk}, {Venter},
  {Venter}, {Welz}, {Williams}, \& {Xaia}}]{2022A&A...657A..56K}
{Knowles}, K., {Cotton}, W.~D., {Rudnick}, L., {et~al.} 2022, \aap, 657, A56

\bibitem[{{Kokotanekov} {et~al.}(2017){Kokotanekov}, {Wise}, {Heald}, {McKean},
  {B{\^\i}rzan}, {Rafferty}, {Godfrey}, {de Vries}, {Intema}, {Broderick},
  {Hardcastle}, {Bonafede}, {Clarke}, {van Weeren}, {R{\"o}ttgering}, {Pizzo},
  {Iacobelli}, {Orr{\'u}}, {Shulevski}, {Riseley}, {Breton},
  {Nikiel-Wroczy{\'n}ski}, {Sridhar}, {Stewart}, {Rowlinson}, {van der Horst},
  {Harwood}, {G{\"u}rkan}, {Carbone}, {Pandey-Pommier}, {Tasse}, {Scaife},
  {Pratley}, {Ferrari}, {Croston}, {Pandey}, {Jurusik}, \&
  {Mulcahy}}]{2017A&A...605A..48K}
{Kokotanekov}, G., {Wise}, M., {Heald}, G.~H., {et~al.} 2017, \aap, 605, A48

\bibitem[{{Lane} {et~al.}(2014){Lane}, {Cotton}, {van Velzen}, {Clarke},
  {Kassim}, {Helmboldt}, {Lazio}, \& {Cohen}}]{2014MNRAS.440..327L}
{Lane}, W.~M., {Cotton}, W.~D., {van Velzen}, S., {et~al.} 2014, \mnras, 440,
  327

\bibitem[{{Lusetti} {et~al.}(2024){Lusetti}, {Bonafede}, {Lovisari}, {Gitti},
  {Ettori}, {Cassano}, {Riseley}, {Govoni}, {Br{\"u}ggen}, {Bruno}, {van
  Weeren}, {Botteon}, {Hoang}, {Gastaldello}, {Ignesti}, {Rossetti}, \&
  {Shimwell}}]{2024A&A...683A.132L}
{Lusetti}, G., {Bonafede}, A., {Lovisari}, L., {et~al.} 2024, \aap, 683, A132

\bibitem[{{Mathews} \& {Brighenti}(2008)}]{2008ApJ...685..128M}
{Mathews}, W.~G. \& {Brighenti}, F. 2008, \apj, 685, 128

\bibitem[{{McNamara} \& {Nulsen}(2007)}]{2007ARA&A..45..117M}
{McNamara}, B.~R. \& {Nulsen}, P.~E.~J. 2007, \araa, 45, 117

\bibitem[{{McNamara} \& {Nulsen}(2012)}]{2012NJPh...14e5023M}
{McNamara}, B.~R. \& {Nulsen}, P.~E.~J. 2012, New Journal of Physics, 14,
  055023

\bibitem[{{McNamara} {et~al.}(2001){McNamara}, {Wise}, {Nulsen}, {David},
  {Carilli}, {Sarazin}, {O'Dea}, {Houck}, {Donahue}, {Baum}, {Voit},
  {O'Connell}, \& {Koekemoer}}]{2001ApJ...562L.149M}
{McNamara}, B.~R., {Wise}, M.~W., {Nulsen}, P.~E.~J., {et~al.} 2001, \apjl,
  562, L149

\bibitem[{{Miley} \& {Perola}(1975)}]{1975A&A....45..223M}
{Miley}, G.~K. \& {Perola}, G.~C. 1975, \aap, 45, 223

\bibitem[{{Miller} \& {Owen}(2001)}]{2001ApJS..134..355M}
{Miller}, N.~A. \& {Owen}, F.~N. 2001, \apjs, 134, 355

\bibitem[{{Morabito} {et~al.}(2016){Morabito}, {Deller}, {R{\"o}ttgering},
  {Miley}, {Varenius}, {Shimwell}, {Mold{\'o}n}, {Jackson}, {Morganti}, {van
  Weeren}, \& {Oonk}}]{2016MNRAS.461.2676M}
{Morabito}, L.~K., {Deller}, A.~T., {R{\"o}ttgering}, H., {et~al.} 2016,
  \mnras, 461, 2676

\bibitem[{{Noordam} \& {de Bruyn}(1982)}]{1982Natur.299..597N}
{Noordam}, J.~E. \& {de Bruyn}, A.~G. 1982, \nat, 299, 597

\bibitem[{{Offringa}(2010)}]{2010ascl.soft10017O}
{Offringa}, A.~R. 2010, {AOFlagger: RFI Software}

\bibitem[{{Offringa}(2016)}]{2016A&A...595A..99O}
{Offringa}, A.~R. 2016, \aap, 595, A99

\bibitem[{{Offringa} {et~al.}(2010){Offringa}, {de Bruyn}, {Biehl}, {Zaroubi},
  {Bernardi}, \& {Pandey}}]{2010MNRAS.405..155O}
{Offringa}, A.~R., {de Bruyn}, A.~G., {Biehl}, M., {et~al.} 2010, \mnras, 405,
  155

\bibitem[{{Offringa} {et~al.}(2014){Offringa}, {McKinley}, {Hurley-Walker},
  {Briggs}, {Wayth}, {Kaplan}, {Bell}, {Feng}, {Neben}, {Hughes}, {Rhee},
  {Murphy}, {Bhat}, {Bernardi}, {Bowman}, {Cappallo}, {Corey}, {Deshpande},
  {Emrich}, {Ewall-Wice}, {Gaensler}, {Goeke}, {Greenhill}, {Hazelton},
  {Hindson}, {Johnston-Hollitt}, {Jacobs}, {Kasper}, {Kratzenberg}, {Lenc},
  {Lonsdale}, {Lynch}, {McWhirter}, {Mitchell}, {Morales}, {Morgan},
  {Kudryavtseva}, {Oberoi}, {Ord}, {Pindor}, {Procopio}, {Prabu}, {Riding},
  {Roshi}, {Shankar}, {Srivani}, {Subrahmanyan}, {Tingay}, {Waterson},
  {Webster}, {Whitney}, {Williams}, \& {Williams}}]{2014MNRAS.444..606O}
{Offringa}, A.~R., {McKinley}, B., {Hurley-Walker}, N., {et~al.} 2014, \mnras,
  444, 606

\bibitem[{{Offringa} \& {Smirnov}(2017)}]{2017MNRAS.471..301O}
{Offringa}, A.~R. \& {Smirnov}, O. 2017, \mnras, 471, 301

\bibitem[{{Pedlar} {et~al.}(1990){Pedlar}, {Ghataure}, {Davies}, {Harrison},
  {Perley}, {Crane}, \& {Unger}}]{1990MNRAS.246..477P}
{Pedlar}, A., {Ghataure}, H.~S., {Davies}, R.~D., {et~al.} 1990, \mnras, 246,
  477

\bibitem[{{Perkins} {et~al.}(2006){Perkins}, {Badran}, {Blaylock}, {Bradbury},
  {Cogan}, {Chow}, {Cui}, {Daniel}, {Falcone}, {Fegan}, {Finley}, {Fortin},
  {Fortson}, {Gillanders}, {Gutierrez}, {Grube}, {Hall}, {Hanna}, {Holder},
  {Horan}, {Hughes}, {Humensky}, {Kenny}, {Kertzman}, {Kieda}, {Kildea},
  {Kosack}, {Krawczynski}, {Krennrich}, {Lang}, {LeBohec}, {Maier}, {Moriarty},
  {Ong}, {Pohl}, {Ragan}, {Rebillot}, {Sembroski}, {Steele}, {Swordy},
  {Valcarcel}, {Vassiliev}, {Wakely}, {Weekes}, {Williams}, \& {VERITAS
  Collaboration}}]{2006ApJ...644..148P}
{Perkins}, J.~S., {Badran}, H.~M., {Blaylock}, G., {et~al.} 2006, \apj, 644,
  148

\bibitem[{{Pfrommer} \& {En{\ss}lin}(2004)}]{2004A&A...413...17P}
{Pfrommer}, C. \& {En{\ss}lin}, T.~A. 2004, \aap, 413, 17

\bibitem[{{Pinzke} \& {Pfrommer}(2010)}]{2010MNRAS.409..449P}
{Pinzke}, A. \& {Pfrommer}, C. 2010, \mnras, 409, 449

\bibitem[{{Planck Collaboration} {et~al.}(2020){Planck Collaboration},
  {Aghanim}, {Akrami}, {Ashdown}, {Aumont}, {Baccigalupi}, {Ballardini},
  {Banday}, {Barreiro}, {Bartolo}, {Basak}, {Battye}, {Benabed}, {Bernard},
  {Bersanelli}, {Bielewicz}, {Bock}, {Bond}, {Borrill}, {Bouchet}, {Boulanger},
  {Bucher}, {Burigana}, {Butler}, {Calabrese}, {Cardoso}, {Carron},
  {Challinor}, {Chiang}, {Chluba}, {Colombo}, {Combet}, {Contreras}, {Crill},
  {Cuttaia}, {de Bernardis}, {de Zotti}, {Delabrouille}, {Delouis}, {Di
  Valentino}, {Diego}, {Dor{\'e}}, {Douspis}, {Ducout}, {Dupac}, {Dusini},
  {Efstathiou}, {Elsner}, {En{\ss}lin}, {Eriksen}, {Fantaye}, {Farhang},
  {Fergusson}, {Fernandez-Cobos}, {Finelli}, {Forastieri}, {Frailis},
  {Fraisse}, {Franceschi}, {Frolov}, {Galeotta}, {Galli}, {Ganga},
  {G{\'e}nova-Santos}, {Gerbino}, {Ghosh}, {Gonz{\'a}lez-Nuevo}, {G{\'o}rski},
  {Gratton}, {Gruppuso}, {Gudmundsson}, {Hamann}, {Handley}, {Hansen},
  {Herranz}, {Hildebrandt}, {Hivon}, {Huang}, {Jaffe}, {Jones}, {Karakci},
  {Keih{\"a}nen}, {Keskitalo}, {Kiiveri}, {Kim}, {Kisner}, {Knox},
  {Krachmalnicoff}, {Kunz}, {Kurki-Suonio}, {Lagache}, {Lamarre}, {Lasenby},
  {Lattanzi}, {Lawrence}, {Le Jeune}, {Lemos}, {Lesgourgues}, {Levrier},
  {Lewis}, {Liguori}, {Lilje}, {Lilley}, {Lindholm}, {L{\'o}pez-Caniego},
  {Lubin}, {Ma}, {Mac{\'\i}as-P{\'e}rez}, {Maggio}, {Maino}, {Mandolesi},
  {Mangilli}, {Marcos-Caballero}, {Maris}, {Martin}, {Martinelli},
  {Mart{\'\i}nez-Gonz{\'a}lez}, {Matarrese}, {Mauri}, {McEwen}, {Meinhold},
  {Melchiorri}, {Mennella}, {Migliaccio}, {Millea}, {Mitra},
  {Miville-Desch{\^e}nes}, {Molinari}, {Montier}, {Morgante}, {Moss}, {Natoli},
  {N{\o}rgaard-Nielsen}, {Pagano}, {Paoletti}, {Partridge}, {Patanchon},
  {Peiris}, {Perrotta}, {Pettorino}, {Piacentini}, {Polastri}, {Polenta},
  {Puget}, {Rachen}, {Reinecke}, {Remazeilles}, {Renzi}, {Rocha}, {Rosset},
  {Roudier}, {Rubi{\~n}o-Mart{\'\i}n}, {Ruiz-Granados}, {Salvati}, {Sandri},
  {Savelainen}, {Scott}, {Shellard}, {Sirignano}, {Sirri}, {Spencer},
  {Sunyaev}, {Suur-Uski}, {Tauber}, {Tavagnacco}, {Tenti}, {Toffolatti},
  {Tomasi}, {Trombetti}, {Valenziano}, {Valiviita}, {Van Tent}, {Vibert},
  {Vielva}, {Villa}, {Vittorio}, {Wandelt}, {Wehus}, {White}, {White},
  {Zacchei}, \& {Zonca}}]{2020A&A...641A...6P}
{Planck Collaboration}, {Aghanim}, N., {Akrami}, Y., {et~al.} 2020, \aap, 641,
  A6

\bibitem[{{Rector} {et~al.}(1999){Rector}, {Stocke}, \&
  {Perlman}}]{1999ApJ...516..145R}
{Rector}, T.~A., {Stocke}, J.~T., \& {Perlman}, E.~S. 1999, \apj, 516, 145

\bibitem[{{Rengelink} {et~al.}(1997){Rengelink}, {Tang}, {de Bruyn}, {Miley},
  {Bremer}, {Roettgering}, \& {Bremer}}]{1997A&AS..124..259R}
{Rengelink}, R.~B., {Tang}, Y., {de Bruyn}, A.~G., {et~al.} 1997, \aaps, 124,
  259

\bibitem[{{Richard-Laferri{\`e}re} {et~al.}(2020){Richard-Laferri{\`e}re},
  {Hlavacek-Larrondo}, {Nemmen}, {Rhea}, {Taylor}, {Prasow-{\'E}mond},
  {Gendron-Marsolais}, {Latulippe}, {Edge}, {Fabian}, {Sanders}, {Hogan}, \&
  {Demontigny}}]{2020MNRAS.499.2934R}
{Richard-Laferri{\`e}re}, A., {Hlavacek-Larrondo}, J., {Nemmen}, R.~S.,
  {et~al.} 2020, \mnras, 499, 2934

\bibitem[{{Riseley} {et~al.}(2024){Riseley}, {Bonafede}, {Bruno}, {Botteon},
  {Rossetti}, {Biava}, {Bonnassieux}, {Loi}, {Vernstrom}, \&
  {Balboni}}]{2024A&A...686A..44R}
{Riseley}, C.~J., {Bonafede}, A., {Bruno}, L., {et~al.} 2024, \aap, 686, A44

\bibitem[{{Rossetti} \& {Molendi}(2010)}]{2010A&A...510A..83R}
{Rossetti}, M. \& {Molendi}, S. 2010, \aap, 510, A83

\bibitem[{{Sanders} {et~al.}(2020){Sanders}, {Dennerl}, {Russell}, {Eckert},
  {Pinto}, {Fabian}, {Walker}, {Tamura}, {ZuHone}, \&
  {Hofmann}}]{2020A&A...633A..42S}
{Sanders}, J.~S., {Dennerl}, K., {Russell}, H.~R., {et~al.} 2020, \aap, 633,
  A42

\bibitem[{{Savini} {et~al.}(2018){Savini}, {Bonafede}, {Br{\"u}ggen}, {van
  Weeren}, {Brunetti}, {Intema}, {Botteon}, {Shimwell}, {Wilber}, {Rafferty},
  {Giacintucci}, {Cassano}, {Cuciti}, {de Gasperin}, {R{\"o}ttgering}, {Hoeft},
  \& {White}}]{2018MNRAS.478.2234S}
{Savini}, F., {Bonafede}, A., {Br{\"u}ggen}, M., {et~al.} 2018, \mnras, 478,
  2234

\bibitem[{{Scaife} \& {Heald}(2012)}]{2012MNRAS.423L..30S}
{Scaife}, A. M.~M. \& {Heald}, G.~H. 2012, \mnras, 423, L30

\bibitem[{{Sijbring}(1993)}]{1993PhDT.......392S}
{Sijbring}, L.~G. 1993, PhD thesis, University of Groningen, Netherlands

\bibitem[{{Smirnov}(2011{\natexlab{a}})}]{2011A&A...527A.106S}
{Smirnov}, O.~M. 2011{\natexlab{a}}, \aap, 527, A106

\bibitem[{{Smirnov}(2011{\natexlab{b}})}]{2011A&A...527A.107S}
{Smirnov}, O.~M. 2011{\natexlab{b}}, \aap, 527, A107

\bibitem[{{Soboleva} {et~al.}(1983){Soboleva}, {Temirova}, {Timofeeva}, \&
  {Aliakberov}}]{1983SvAL....9..305S}
{Soboleva}, N.~S., {Temirova}, A.~V., {Timofeeva}, G.~M., \& {Aliakberov},
  K.~D. 1983, Soviet Astronomy Letters, 9, 305

\bibitem[{{Thierbach} {et~al.}(2003){Thierbach}, {Klein}, \&
  {Wielebinski}}]{2003A&A...397...53T}
{Thierbach}, M., {Klein}, U., \& {Wielebinski}, R. 2003, \aap, 397, 53

\bibitem[{{Tully} {et~al.}(2016){Tully}, {Courtois}, \&
  {Sorce}}]{2016AJ....152...50T}
{Tully}, R.~B., {Courtois}, H.~M., \& {Sorce}, J.~G. 2016, \aj, 152, 50

\bibitem[{{van der Tol} {et~al.}(2007){van der Tol}, {Jeffs}, \& {van der
  Veen}}]{2007ITSP...55.4497V}
{van der Tol}, S., {Jeffs}, B.~D., \& {van der Veen}, A.~J. 2007, IEEE
  Transactions on Signal Processing, 55, 4497

\bibitem[{{van Haarlem} {et~al.}(2013){van Haarlem}, {Wise}, {Gunst}, {Heald},
  {McKean}, {Hessels}, {de Bruyn}, {Nijboer}, {Swinbank}, {Fallows},
  {Brentjens}, {Nelles}, {Beck}, {Falcke}, {Fender}, {H{\"o}randel},
  {Koopmans}, {Mann}, {Miley}, {R{\"o}ttgering}, {Stappers}, {Wijers},
  {Zaroubi}, {van den Akker}, {Alexov}, {Anderson}, {Anderson}, {van Ardenne},
  {Arts}, {Asgekar}, {Avruch}, {Batejat}, {B{\"a}hren}, {Bell}, {Bell}, {van
  Bemmel}, {Bennema}, {Bentum}, {Bernardi}, {Best}, {B{\^\i}rzan}, {Bonafede},
  {Boonstra}, {Braun}, {Bregman}, {Breitling}, {van de Brink}, {Broderick},
  {Broekema}, {Brouw}, {Br{\"u}ggen}, {Butcher}, {van Cappellen}, {Ciardi},
  {Coenen}, {Conway}, {Coolen}, {Corstanje}, {Damstra}, {Davies}, {Deller},
  {Dettmar}, {van Diepen}, {Dijkstra}, {Donker}, {Doorduin}, {Dromer}, {Drost},
  {van Duin}, {Eisl{\"o}ffel}, {van Enst}, {Ferrari}, {Frieswijk}, {Gankema},
  {Garrett}, {de Gasperin}, {Gerbers}, {de Geus}, {Grie{\ss}meier}, {Grit},
  {Gruppen}, {Hamaker}, {Hassall}, {Hoeft}, {Holties}, {Horneffer}, {van der
  Horst}, {van Houwelingen}, {Huijgen}, {Iacobelli}, {Intema}, {Jackson},
  {Jelic}, {de Jong}, {Juette}, {Kant}, {Karastergiou}, {Koers}, {Kollen},
  {Kondratiev}, {Kooistra}, {Koopman}, {Koster}, {Kuniyoshi}, {Kramer},
  {Kuper}, {Lambropoulos}, {Law}, {van Leeuwen}, {Lemaitre}, {Loose}, {Maat},
  {Macario}, {Markoff}, {Masters}, {McFadden}, {McKay-Bukowski}, {Meijering},
  {Meulman}, {Mevius}, {Middelberg}, {Millenaar}, {Miller-Jones}, {Mohan},
  {Mol}, {Morawietz}, {Morganti}, {Mulcahy}, {Mulder}, {Munk}, {Nieuwenhuis},
  {van Nieuwpoort}, {Noordam}, {Norden}, {Noutsos}, {Offringa}, {Olofsson},
  {Omar}, {Orr{\'u}}, {Overeem}, {Paas}, {Pandey-Pommier}, {Pandey}, {Pizzo},
  {Polatidis}, {Rafferty}, {Rawlings}, {Reich}, {de Reijer}, {Reitsma},
  {Renting}, {Riemers}, {Rol}, {Romein}, {Roosjen}, {Ruiter}, {Scaife}, {van
  der Schaaf}, {Scheers}, {Schellart}, {Schoenmakers}, {Schoonderbeek},
  {Serylak}, {Shulevski}, {Sluman}, {Smirnov}, {Sobey}, {Spreeuw}, {Steinmetz},
  {Sterks}, {Stiepel}, {Stuurwold}, {Tagger}, {Tang}, {Tasse}, {Thomas},
  {Thoudam}, {Toribio}, {van der Tol}, {Usov}, {van Veelen}, {van der Veen},
  {ter Veen}, {Verbiest}, {Vermeulen}, {Vermaas}, {Vocks}, {Vogt}, {de Vos},
  {van der Wal}, {van Weeren}, {Weggemans}, {Weltevrede}, {White}, {Wijnholds},
  {Wilhelmsson}, {Wucknitz}, {Yatawatta}, {Zarka}, {Zensus}, \& {van
  Zwieten}}]{2013A&A...556A...2V}
{van Haarlem}, M.~P., {Wise}, M.~W., {Gunst}, A.~W., {et~al.} 2013, \aap, 556,
  A2

\bibitem[{{van Weeren} {et~al.}(2019){van Weeren}, {de Gasperin}, {Akamatsu},
  {Br{\"u}ggen}, {Feretti}, {Kang}, {Stroe}, \&
  {Zandanel}}]{2019SSRv..215...16V}
{van Weeren}, R.~J., {de Gasperin}, F., {Akamatsu}, H., {et~al.} 2019, \ssr,
  215, 16

\bibitem[{{van Weeren} {et~al.}(2021){van Weeren}, {Shimwell}, {Botteon},
  {Brunetti}, {Br{\"u}ggen}, {Boxelaar}, {Cassano}, {Di Gennaro},
  {Andrade-Santos}, {Bonnassieux}, {Bonafede}, {Cuciti}, {Dallacasa}, {de
  Gasperin}, {Gastaldello}, {Hardcastle}, {Hoeft}, {Kraft}, {Mandal},
  {Rossetti}, {R{\"o}ttgering}, {Tasse}, \& {Wilber}}]{2021A&A...651A.115V}
{van Weeren}, R.~J., {Shimwell}, T.~W., {Botteon}, A., {et~al.} 2021, \aap,
  651, A115

\bibitem[{{van Weeren} {et~al.}(2024){van Weeren}, {Timmerman}, {Vaidya},
  {Gendron-Marsolais}, {Botteon}, {Roberts}, {Hlavacek-Larrondo}, {Bonafede},
  {Br{\"u}ggen}, {Brunetti}, {Cassano}, {Cuciti}, {Edge}, {Gastaldello},
  {Groeneveld}, \& {Shimwell}}]{2024A&A...692A..12V}
{van Weeren}, R.~J., {Timmerman}, R., {Vaidya}, V., {et~al.} 2024, \aap, 692,
  A12

\bibitem[{{Venturi} {et~al.}(2017){Venturi}, {Rossetti}, {Brunetti},
  {Farnsworth}, {Gastaldello}, {Giacintucci}, {Lal}, {Rudnick}, {Shimwell},
  {Eckert}, {Molendi}, \& {Owers}}]{2017A&A...603A.125V}
{Venturi}, T., {Rossetti}, M., {Brunetti}, G., {et~al.} 2017, \aap, 603, A125

\bibitem[{{Wise} {et~al.}(2007){Wise}, {McNamara}, {Nulsen}, {Houck}, \&
  {David}}]{2007ApJ...659.1153W}
{Wise}, M.~W., {McNamara}, B.~R., {Nulsen}, P.~E.~J., {Houck}, J.~C., \&
  {David}, L.~P. 2007, \apj, 659, 1153

\bibitem[{{Zheng} {et~al.}(2017){Zheng}, {Tegmark}, {Dillon}, {Kim}, {Liu},
  {Neben}, {Jonas}, {Reich}, \& {Reich}}]{2017MNRAS.464.3486Z}
{Zheng}, H., {Tegmark}, M., {Dillon}, J.~S., {et~al.} 2017, \mnras, 464, 3486

\bibitem[{{ZuHone} {et~al.}(2013){ZuHone}, {Markevitch}, {Brunetti}, \&
  {Giacintucci}}]{2013ApJ...762...78Z}
{ZuHone}, J.~A., {Markevitch}, M., {Brunetti}, G., \& {Giacintucci}, S. 2013,
  \apj, 762, 78

\end{thebibliography}

\begin{appendix}
\onecolumn

\FloatBarrier

\section{Masked regions}

\begin{figure}[ht]
    \centering
    \includegraphics[width=0.7\linewidth]{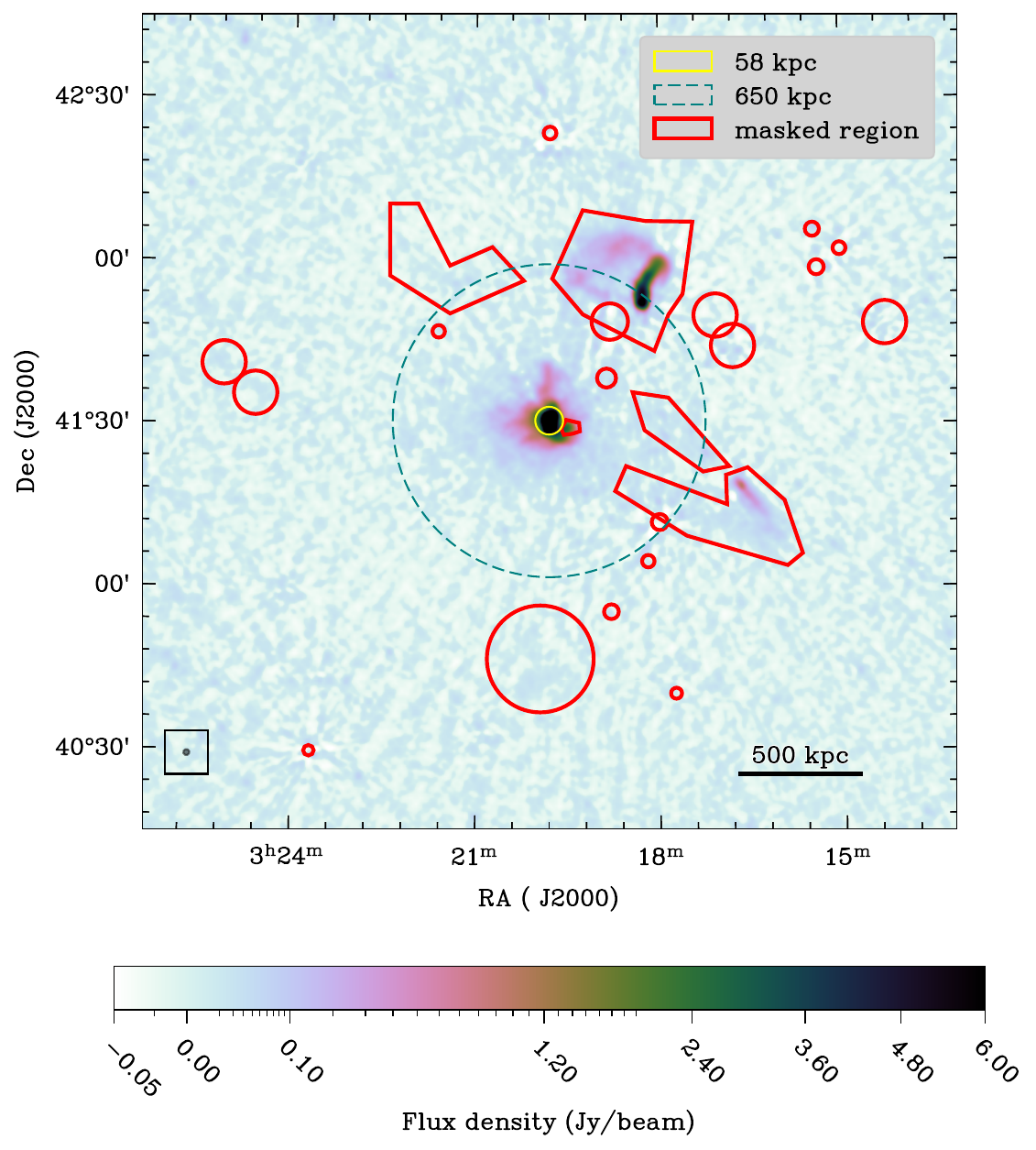}
    \caption{LOFAR LBA image of the Perseus cluster tapered to 68\arcsec{} and with compact sources subtracted. The red areas indicate regions that were excluded due to the presence of contaminating sources, for constructing the radial radio surface brightness profile. The yellow circle (154\arcsec{} radius, 58~kpc) indicates the region that is affected by the presence of 3C\,84. The double-exponential fit for radio surface brightness profile was performed on the data between the yellow circle and the dashed teal circle at 650~kpc radius.}
    \label{fig:masked}
\end{figure}

\FloatBarrier
\newpage

\section{Flux density reference sources}

\begin{figure*}[ht]
    \centering
    \includegraphics[width=0.9\linewidth]{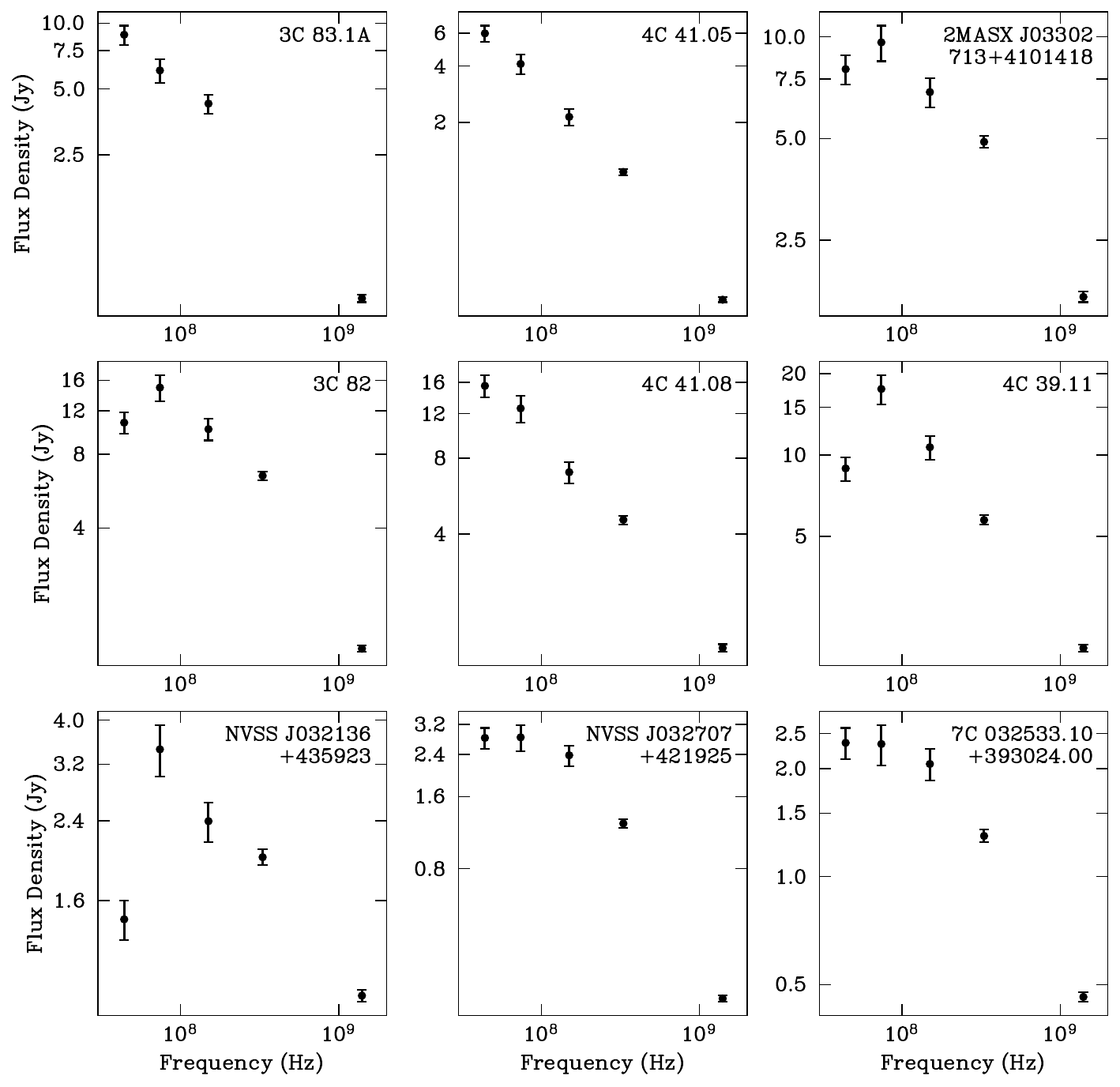}
    \caption{Integrated flux densities of nine bright sources between 44~MHz and 1.4~GHz in the field of view of the LOFAR observation of the Perseus cluster. The sources in this figure are marked with a circle in Fig. \ref{fig:finder_plot}. The lowest frequency point is the flux density measured in this work. The other surveys used in this figure are (in order of increasing frequency) are: VLSSr \citep[74 MHz;][]{2014MNRAS.440..327L}; TGSS ADR1 \citep[150 MHz;][]{2017A&A...598A..78I}, WENSS \citep[330 MHz;][]{1997A&AS..124..259R} and, NVSS \citep[1.4 GHz;][]{1998AJ....115.1693C}.}
    \label{fig:flux_check}
\end{figure*}  
\begin{figure*}
    \centering
    \includegraphics[width=0.95\linewidth]{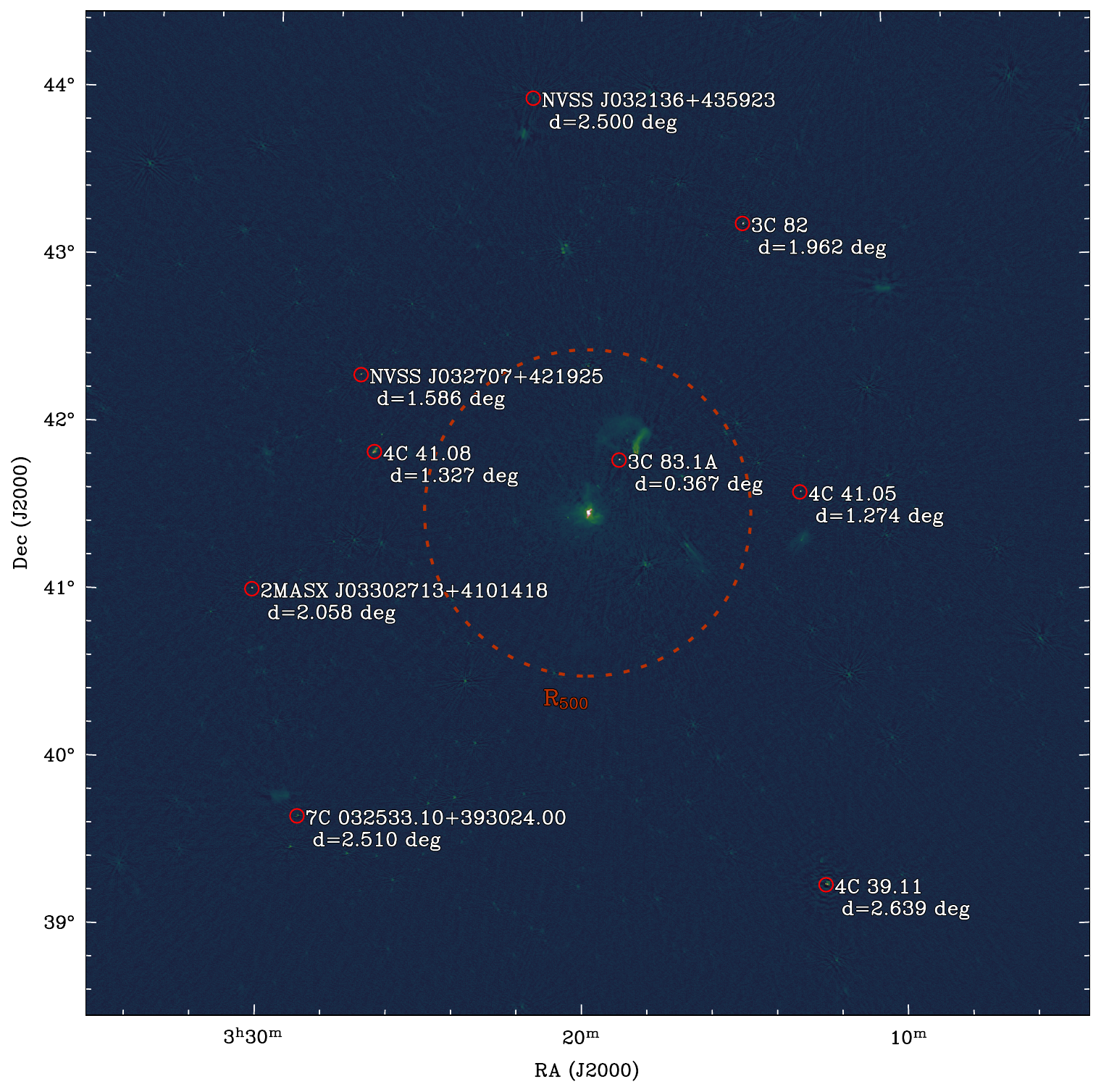}
    \caption{Wide-field image of our observation of the Perseus cluster. The marked sources are used in Fig.~\ref{fig:flux_check}  to check the accuracy of the flux density scale. The angular distance ($d$) to the Perseus cluster (i.e. pointing) centre is indicated near each source on the image. The dotted red circle indicates the extend of the $R_{500}$ of the Perseus cluster. The giant radio halo extends a bit beyond half the $R_{500}$, so we do not expect any flux scale issues due to direction dependent effects.}
    \label{fig:finder_plot}
\end{figure*}

\FloatBarrier
\clearpage

\section{Spectral index error maps}

\begin{figure*}[ht]
    \centering
    \includegraphics[width=0.45\linewidth]{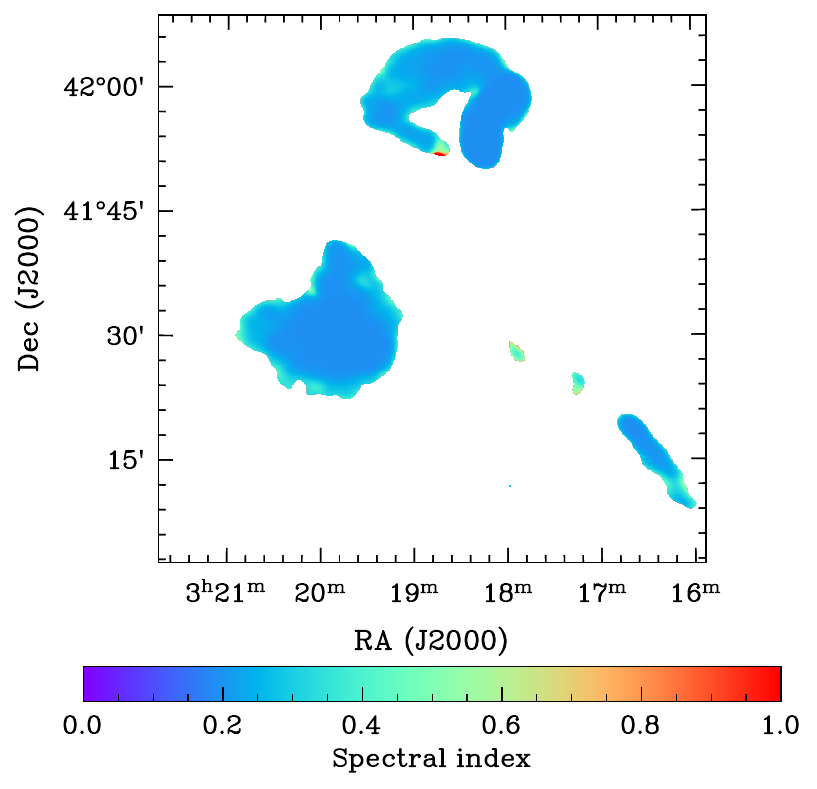}
    \includegraphics[width=0.45\linewidth]{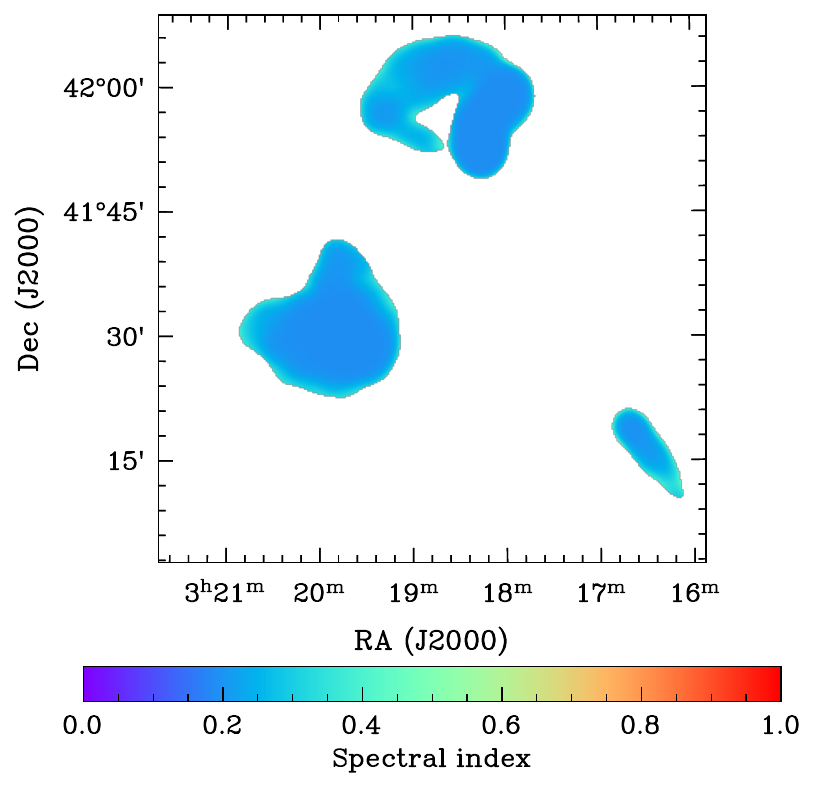} 
    \caption{Spectral index uncertainty maps corresponding to spectral index maps presented in Fig. \ref{fig:spix_map_large} at 60\arcsec (\textit{left}) and  120\arcsec (\textit{right}) resolution.}
    \label{fig:spixerr}
\end{figure*}

\begin{figure*}[ht]
    \centering
    \includegraphics[width=0.5\linewidth]{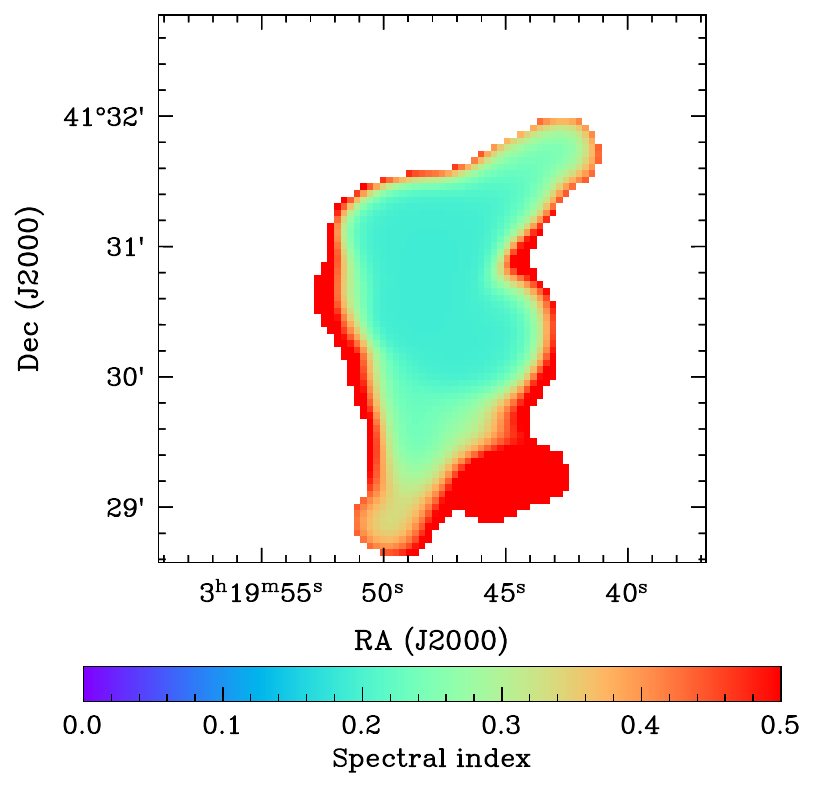}
    \caption{Spectral index uncertainty map corresponding to spectral index map presented in Fig. \ref{fig:zoominspix}.}
    \label{fig:spix_err}
\end{figure*}

\FloatBarrier
\newpage

\section{Cutouts of tailed radio galaxies}

\begin{figure*}[!ht]
    \centering
    \includegraphics[width=\textwidth]{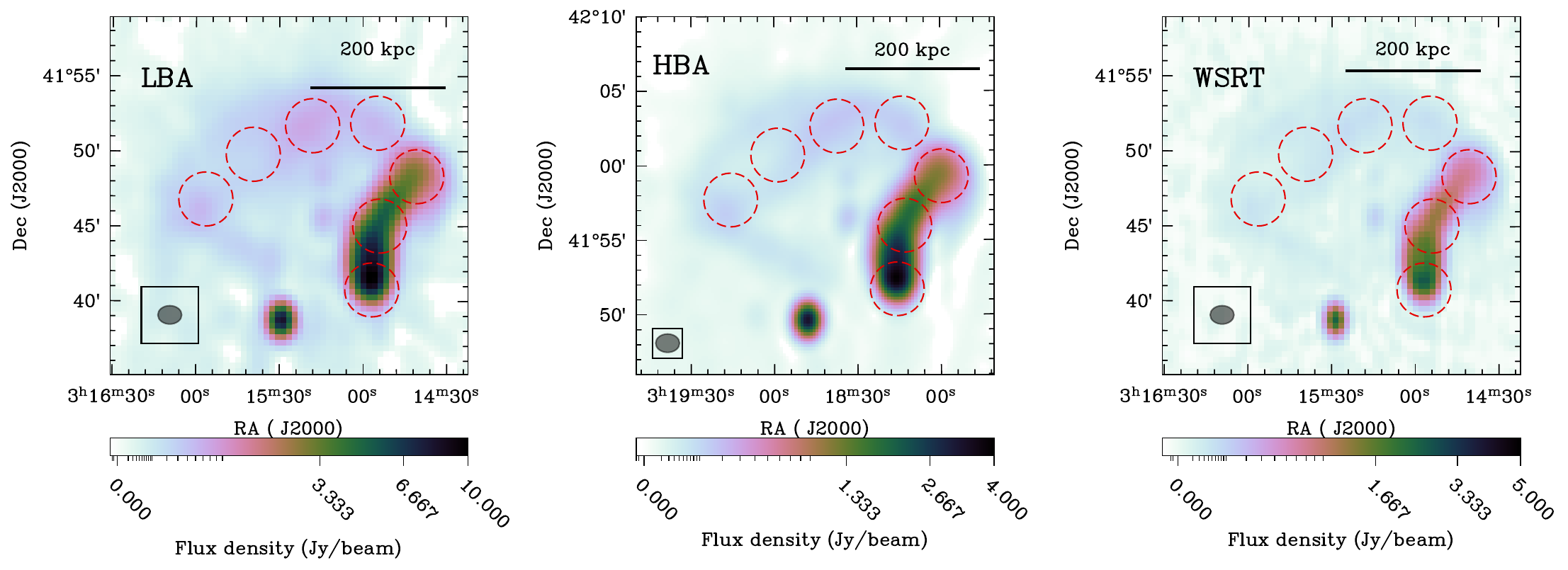}
    \caption{Cutout of NGC~1265. The circles indicate the regions used for measuring the spectral index along the tails in Fig. \ref{fig: profiles}.}
    \label{fig: ngcthrice}
\end{figure*}

\begin{figure*}[!ht]
    \centering
    \includegraphics[width=\textwidth]{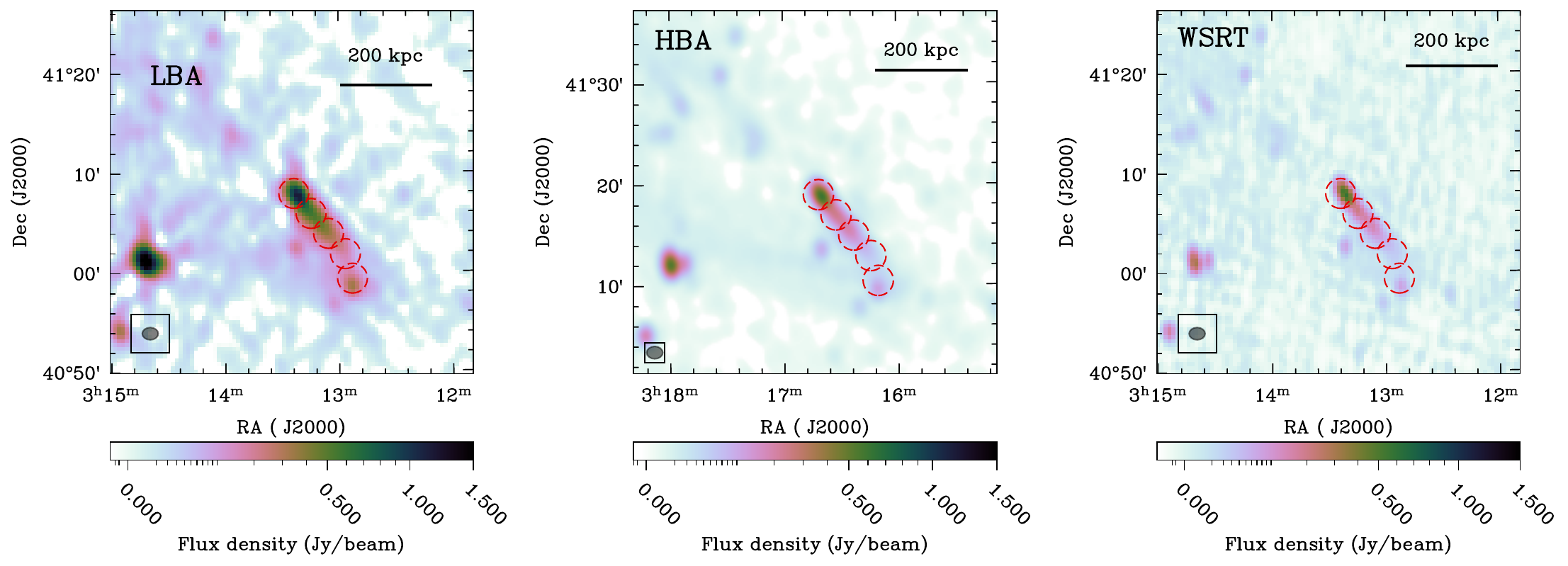}
    \caption{Cutout of IC~310. The circles indicate the regions used for measuring the spectral index along the tails in Fig. \ref{fig: profiles}}
    \label{fig: icthrice}
\end{figure*}

\end{appendix}

\end{document}